 \documentclass[10pt,preprint]{aastex}  
 \usepackage{natbib}
\def\ang{\AA}

\def\gapprox{\lower.4ex\hbox{$\;\buildrel >\over{\scriptstyle\sim}\;$}}
\def\lapprox{\lower.4ex\hbox{$\;\buildrel <\over{\scriptstyle\sim}\;$}}

\shortauthors{ASCHWANDEN ET AL. 2015}
\shorttitle{Global Energetics of Solar Flares. IV.}

\begin{document}

\title{         Global Energetics of Solar Flares: 
		IV. Coronal Mass Ejection Energetics }

\author{        Markus J. Aschwanden$^1$}

\affil{		$^1)$ Lockheed Martin, 
		Solar and Astrophysics Laboratory, 
                Org. A021S, Bldg.~252, 3251 Hanover St.,
                Palo Alto, CA 94304, USA;
                e-mail: aschwanden@lmsal.com}

\begin{abstract}
This study entails the fourth part of a global flare energetics project,
in which the mass $m_{\mathrm{cme}}$, kinetic energy $E_{\mathrm{kin}}$, 
and the gravitational potential energy $E_{\mathrm{grav}}$ of coronal 
mass ejections (CMEs) is measured in 399 M and X-class 
flare events observed during the first 3.5 yrs of the {\sl Solar Dynamics 
Observatory (SDO)} mission, using a new method based on the EUV dimming effect. 
The EUV dimming is modeled in terms of a radial adiabatic expansion process,
which is fitted to the observed evolution of the total emission measure
of the CME source region. The model derives the evolution of the mean 
electron density, the emission measure, the bulk plasma expansion velocity, 
the mass, and the energy in the CME source region.  The EUV dimming method 
is truly complementary to the Thomson scattering method in white light, 
which probes the CME evolution in the heliosphere
at $r \gapprox 2 R_{\odot}$, while the EUV dimming method tracks the CME
launch in the corona. We compare the CME parameters obtained in white light
with the LASCO/C2 coronagraph with those obtained from EUV dimming with
the {\sl Atmospheric Imaging Assembly (AIA)} onboard SDO for all identical
events in both data sets. We investigate correlations between CME parameters,
the relative timing with flare parameters, frequency occurrence distributions,
and the energy partition between magnetic, thermal, nonthermal, and CME energies.
CME energies are found to be systematically lower than the dissipated
magnetic energies, which is consistent with a magnetic origin of CMEs. 
\end{abstract}

\section{		    INTRODUCTION			}

We undertake a systematic survey of the global energetics of solar flares and
coronal mass ejections (CME) observed during the SDO era, which includes
all M and X-class flares during the first 3.5 years of the SDO mission,
covering 399 flare events.  This project embodies the most comprehensive
survey of the various types of energies that can be detected during flares
and CME events, including the dissipated magnetic energy (Aschwanden Xu, 
and Jing 2014; Paper I), the multi-thermal energy (Aschwanden et al.~2015; 
Paper II), the nonthermal energy (Aschwanden et al.~2016; Paper III), 
and the kinetic and gravitational energy of associated CMEs, which is
the focus of this study. The kinetic and gravitational CME energies are
calculated here with a new method based on the extreme ultra-violet (EUV)
dimming effect, using 
{\sl Atmospheric Imaging Assembly (AIA)} data (Lemen et al.~2012) onboard the 
{\sl Solar Dynamics Observatory (SDO)} (Pesnell et al.~2011), and are 
compared with data from the previously established CME catalog based on 
the white-light method (courtesy of Nat Gopalswamy and Seiji Yashiro), 
using observations from the {\sl Large-Angle and Spectrometric Coronagraph 
Experiment (LASCO)} onboard the {\sl Solar and Heliospheric Observatory 
(SOHO)} (Brueckner et al.~1995). The basic difference between the 
EUV and the white-light method of CME detection is pictured in Fig.~1. 
Both are remote-sensing methods, but one is probing the coronal CME 
source region (via EUV dimming) while the other detects scattered 
white light at a distance of a few solar radii, outside the edge
of the coronagraph occulter disk.

While the speeds, masses, and kinetic energies of CMEs were previously
measured with coronagraphs (Jackson and Hildner 1978; Howard et al.~1985;
Vourlidas et al.~2000, 2002; Subramanian and Vourlidas 2007; Colaninno
and Vourlidas 2009), based on the method of Thomson scattering of white light
off CME particles (Minnaert 1930; van de Hulst 1950; Billings 1966; 
Vourlidas and Howard 2006), an alternative model is developed here based on the 
extreme ultra-violet (EUV)
dimming effect in the coronal CME source region. The idea of interpreting
EUV dimming in terms of mass depletion due to rapid expansion during the
CME launch and its observational evidence in soft X-rays and EUV has been 
discussed in a number of previous studies (Hudson et al.~1996;
Sterling and Hudson 1997; Thompson et al.~1998, 2000; Zarro et al.~1999; 
Harrison and Lyons 2000;
Sterling et al.~2000; Harra and Sterling 2001; Harrison et al.~2003; 
Howard and Harrison 2004; Zhukov and Auchere 2004; Harra et al. 2007;
Harrison and Bewsher 2007; Moore and Sterling 2007; Bewsher et al.~2008; 
Attrill et al.~2008; Reinard and Biesecker 2008, 2009; Robbrecht et al.~2009; 
Aschwanden and W\"ulser 2011; Woods et al.~2011, 2012; Tian et al.~2012; 
Cheng et al.~2012; Krista and Reinard 2013; Howard and Harrison 2013; 
Nieves-Chinchilla et al.~2013; Bein et al.~2013; Mason et al.~2014; 
Woods 2014; Cheng and Qiu 2016), mostly in a qualitative manner, while 
we fit a quantitative EUV dimming model to the data here, for the first time, 
with ample statistics. The principle of measuring CME masses with the EUV 
dimming method in the coronal CME source region has been quantitatively 
tested with data from the {\sl Extreme Ultraviolet Imager (EUVI)}
onboard the {\sl Solar-Terrestrial Relationship Observatory (STEREO)}
(Kaiser et al.~2008), demonstrating 
agreement of the CME masses within $m_A/m_B = 1.3\pm0.6$ between the two STEREO 
spacecraft, and within $m_{\mathrm{EUVI}}/m_{\mathrm{COR2}} = 1.1 \pm 0.3$ 
between the STEREO/EUVI 
spacecraft and the white-light coronagraph STEREO/COR2, for a small sample of CME 
events (Aschwanden et al.~2009). In this study here we develop the EUV dimming 
method further, so that it allows us (besides the CME mass and gravitational energy) 
also to determine the speed and kinetic energy of CMEs. We model the dynamics 
of EUV dimming during the launch of a CME in the lower corona with a model of 
radial adiabatic expansion, which predicts the temporal dimming profile as a 
function of the acceleration rate, and can be fitted to the data to 
obtain the acceleration rate, the speed, and the height-time profile of CMEs. 
This quantitative model permits us also to predict the expected detection time 
and CME velocity measured in the C2 or C3 field-of-view of the LASCO/SOHO coronagraph.

In Section 2 and Appendix A we derive the theoretical model of EUV dimming from 
first principles,
while the counterpart model based on Thomson scattering with white-light data
is briefly summarized in Appendix B. Data analysis and results of 399 flare/CME 
events observed with AIA/SDO is presented in Section 3, along with white-light
measurements of identical events observed with LASCO/SOHO.
A discussion of the various aspects of the two methods in EUV and white-light
is given in Section 4, with conclusions in Section 5. 

\section{	THE EUV DIMMING METHOD 				}

Our ultimate goal is to understand the kinematics, dynamics, and energetics
of CMEs, in particular in the context of the global energetics of solar flare 
events. For this purpose we perform quantitative measurements of physical
parameters in CMEs with the EUV dimming method, which includes the
geometric size $L$, area $A$, and volume $V$ of the dimming region, the mean 
electron density $n_e(t)$, the electron temperature $T_e(t)$, the differential emission 
measure distribution $dEM(T,t)/dT$, the total mass $m_{\mathrm{cme}}$, the acceleration $a(t)$, 
the velocity $v(t)$, the propagation distance from Sun center $x(t)$, 
the kinetic $E_{\mathrm{kin}}$, and the gravitational energy $E_{\mathrm{grav}}$. 
We develop here, for the first time, a quantitative physical model that is based 
on the radial adiabatic expansion process and can be fitted to the observed 
EUV dimming time profiles $EM(t)$.

\subsection{	The Data Setup 				 	}

The primary data set for this study consists of all solar GOES M and X-class flare
events observed with AIA/SDO during the first 3.5 years of the SDO mission 
(June 2010 - Jan 2014), which amounts to 399 events.
For each flare event, the start time $t_{\mathrm{start}}$, the peak time
$t_{\mathrm{peak}}$, and end time $t_{\mathrm{end}}$ are extracted from the NOAA/GOES flare catalog.
For the analysis of the EUV dimming profiles, we include a time margin of
$\Delta t_{\mathrm{margin}} = 30$ min after the flare end, yielding a time window of
$[t_{\mathrm{start}},t_{\mathrm{end}}+\Delta t_{\mathrm{margin}}]$ for each event, during which we analyze
AIA data with a cadence of $\Delta t = 2$ min, or shorter if necessary,
amounting to $n_t \approx 10-200$ time frames per event.
We are using EUV images from SDO/AIA in all 6 coronal wavelengths, i.e., in
94, 131, 171, 193, 211, 335 \ang . We select a field-of-view of $FOV=0.5$
$R_{\odot}$ centered at the heliographic position of the flare site
(obtained from the {\sl SolarSoft (SSW)} latest events archive at 
{\sl http://www.lmsal.com/solarsoft/latest$\_$events$\_$archive.html},
courtesy of Sam Freeland). The size of the field-of-view covers
the initial expansion of the CME and the associated dimming region within
a half solar radius, which is found to be sufficient to model the early
phase of the CME origination where EUV data fitting is sensitive.
Since the CME evolution is traced from 0.1 to 7.0 hours in various 
events, we correct for the solar rotation in the extraction of each time 
sequence of subimages, so that the chosen field-of-view tracks the 
rotating heliographic position of the flare, CME, and dimming region.

\subsection{	The Differential Emission Measure (DEM) of CMEs 	}

Previous analysis of EUV dimming during CMEs has been performed on
soft X-ray or EUV images and thus is strongly wavelength-dependent.
The set of coronal images obtained with AIA/SDO has the great advantage
that essentially the entire relevant temperature range observed in the
solar corona and in flares is covered, in a nominal range of
$T_e \approx 0.5 - 20$ MK. The DEM distribution of CMEs has been measured
in a few previous studies only, using 3 temperature filters from EUVI/STEREO 
(Aschwanden et al.~2009a), or 6 filters from AIA (Cheng et al.~2012).
Reconstructing a DEM distribution over the full coronal temperature range 
allows us to calculate the total mass of a CME in an wavelength-independent
manner.

In order to determine the {\sl differential emission measure
(DEM)} distribution we apply the spatial-synthesis method, which fits
single-Gaussian DEM functions to the EUV fluxes in each macro-pixel, 
\begin{equation}
        {\mathrm{d}EM(T,x,y) \over \mathrm{d}T}= EM_{p}(x,y)
                \exp{\left(- {[\log(T)-\log(T_{p}(x,y))]^2
                \over 2 \sigma_T^2(x,y) } \right)} \ ,
\end{equation}
characterized by the three parameters $[EM_{p}$, $T_{p}$, 
$\sigma_{p}$] for each macro-pixel position $[x,y]$,
where we choose a size of $32 \times 32$ pixels for a macro-pixel, 
which was found to be a good trade-off between computational speed and 
the spatial variation of thermal structures. In other words, using
smaller macro-pixel sizes did not increase the accuracy of the
total emission measure. The total DEM distribution function sampled over 
the entire field-of-view in the EUV dimming region of the CME is 
then obtained by integrating over the full temperature range and by
summing over all macro-pixels,
\begin{equation}
	EM_{\mathrm{tot}}=\int \int \int {\mathrm{d}EM(T,x,y) \over \mathrm{d}T} 
	\ dT\ dx\ dy \ .
\end{equation}

This spatial-synthesis DEM code (Aschwanden et al.~2013) was cross-compared 
with ten other DEM codes in a benchmark test of differential emission measure
codes and multi-thermal energies in solar active regions (Aschwanden et al.~2015)
and was found to be among the three most accurate DEM codes tested therein.
A similar DEM code based on Gaussian basis functions in each image pixel has
recently been developed by the AIA Team also (Cheung et al.~2015).

The spatial-synthesis DEM distributions obtained during a flare event
generally shows a low-temperature peak at $T_e \approx 1.5$ MK 
in the preflare phase of the EUV dimming region. This temperature
serves to estimate the initial density scale height $\lambda$ 
in the dimming region,
\begin{equation}
	\lambda \approx 50\ {\rm Mm} \times \left( {T_e \over 1\ {\rm MK}} \right) \ .
\end{equation}
which yields a density scale height of $\lambda \approx 75$ Mm for a typical
preflare temperature of $T_e \approx 1.5$ MK in the EUV dimming region.

\subsection{	The Geometry of CMEs 	}

The only spatial information on the geometry of the CME available in EUV images 
is the projected area $A_p$ of the EUV dimming region, which defines the CME footpoint 
area.  We determine this area $A_p$ of the dimming region by combining all
macro-pixels that exhibit a significant emission measure decrease 
$\Delta EM_{\mathrm{tot}}$ between the time $t_{\mathrm{max}}$ 
of the maximum emission measure 
$EM_{\mathrm{max}}$ (before the EUV dimming starts) and the time $t_{\mathrm{min}}$ of the minimum
emission measure $E_{\mathrm{min}}$ afterwards (during the EUV dimming phase),
\begin{equation}
	\Delta EM_{\mathrm{tot}} 
	= EM_{\mathrm{max}} - EM_{\mathrm{min}} 
	= EM_{\mathrm{tot}}(t=t_{\mathrm{max}}) 
	- EM_{\mathrm{tot}}(t=t_{\mathrm{min}}) \ .
\end{equation}
In a gravitationally stratified atmosphere, the initial volume $V$ of the
dimming region corresponds to the product of the (unprojected) solar surface 
area $A$ and the vertical scale height $\lambda$,
\begin{equation}
	V = A \lambda = L^2 \lambda \ ,
\end{equation}
if we define a spatial length scale $L$ by the square root of the area $A$,
\begin{equation}
	A = L^2 \ .
\end{equation}	
The dimming area can have any arbitrary 2D shape, because the so-defined
length scale $L$ is just the equivalent linear size of a square-sized area. 
The measurement of the dimming area by summing over all dimmed pixels in the
plane-of-sky, however, yields a projected area $A_p$ that is foreshortened
in the center-to-limb direction, 
\begin{equation}
	A_p = L \ L_p \ ,
\end{equation}
where $L_p$ is the projected length scale (in center-to-limb direction),
while $L$ is the unprojected length scale (in perpendicular direction). 
From the geometric diagram in Fig.~2
we can see that the projected length scale $L_p$ is related to the unprojected
length scale $L$ and vertical scale height $\lambda$ by
\begin{equation}
	L_p =  L \cos{(\rho)} + \lambda \sin{(\rho)} = {A_p \over L}\ ,
\end{equation}
where $\rho$ is the angle of the heliographic position from disk center,
which is according to spherical geometry,
\begin{equation}
	\cos{(\rho)} = cos{(l)} \cos{(b)} \ ,
\end{equation} 
with $l$ being the heliographic longitude and $b$ the heliographic latitude.
We see that the projected length has the limits of $L_p = L$ at disk center
($\rho = 0$), and $L_p = \lambda$ at the limb ($\rho=90^\circ$), respectively.
Inserting $L_p=A_p/L$ from Eq.~(7) into Eq.~(8) leads to a quadratic equation 
of the length scale $L$, which has the algebraic solution,
\begin{equation}
	L = {-\lambda\ \sin{(\rho)} \pm \sqrt{\lambda^2 \sin^2{(\rho)}
	+ 4 A_p \cos{(\rho)}} \over 2 \cos{(\rho)} }.
\end{equation}
With this analytical solution for $L$ we can directly calculate the
unprojected area $A$ (Eq.~6) or volume $V$ (Eq.~5) from the heliographic
position $(l,b)$, the projected area $A_p$ (Eq.~7), and the density
scale height $\lambda$ (Eq.~3). It is worthwhile to use this exact solution,
because the generally used approximation $L \approx \sqrt{A_p}$
overestimates the volume by a factor of $\approx L/\lambda \approx 3$
for typical solar conditions, which propagates as a similar error into 
the estimates of the electron density, mass, kinetic, and gravitational
energy of the CME.

\subsection{The Density and Mass of CMEs }

The total emission measure $EM_{\mathrm{tot}}$ of the dimming region is related to the mean
electron density $n_e$ in a gravitationally stratified atmosphere by,
\begin{equation}
        EM_{\mathrm{tot}} = \int n_e^2({\bf x}) \ dV = n_e^2 V = n_e^2 L^2 \lambda  \ ,
\end{equation}
which yields an explicit expression for the mean electron density $n_e$
in the CME source region, 
\begin{equation}
	n_e = \sqrt{ {EM_{\mathrm{tot}} \over L^2 \lambda } } \ .
\end{equation}
Thus, by measuring the total emission measure $EM_{\mathrm{tot}}$ and the
(projected) dimming area $A_p$ at the heliographic location $(l,b)$, 
and using the density scale height $\lambda$ (Eq.~3),
we can derive the center-to-limb angle $\rho$ (Eq.~9), the unprojected
length scale $L$ (Eq.~10), and the mean electron density $n_e$ (Eq.~12).

Armed with the measurements of the mean electron density $n_e$ and the dimming
volume $V=L^2 \lambda$, we can then directly calculate the mass of the CME, 
evaluated in the CME footpoint area at the time of the CME launch, 
\begin{equation}
        m_{\mathrm{cme}} = n_e\ m_p\ V = n_e\ m_p\ L^2 \lambda \ .
\end{equation}
Typically we find a density of $n_e \approx 10^9$ cm$^{-3}$, a length scale
of $L \approx 10^{10}$ cm, a scale height of $\lambda \approx 7.5 \times 10^9$
cm in the coronal dimming regions, which yields a typical CME mass of
$m_{\mathrm{cme}} \approx 1.3 \times 10^{15}$ g. 

\subsection{The Kinematics of CMEs }

White-light coronagraph observations provide CME speeds at a distance of $x > 2.2
R_{\odot}$ (for LASCO/C2) from Sun center, which is often close to a constant value,
but the acceleration profile in the CME source region in the lower corona is unknown. 
Time sequences of AIA images, in contrast, provide kinematic parameters of acceleration
$a(t)$, speed $v(t)$, and height $x(t)$ of a CME all the way back to the initial
origin of the CME. We attempt to measure these kinematic parameters by forward-fitting
a kinematic model to the observed EUV dimming profile $EM(t)$. The simplest
kinematic model contains a short initial acceleration time $\tau_a$ that is temporally
not resolved with our typical time cadence of $\Delta t=2$ min. Since the altitude range of
significant CME acceleration is estimated to be confined to $h \lapprox 0.1 R_{\odot}$, 
the corresponding acceleration time interval is $\tau_a \approx 2$ min for a CME with 
a typical final speed of $v \approx 1000$ km s$^{-1}$, so that the details of the 
acceleration time profile can be neglected for data fitting with a similar time 
resolution. Therefore, we set the acceleration time equal to the time resolution, 
i.e., $\tau_a = \Delta t$, while the acceleration is zero before and afterwards. 

However, for detailed modeling of the acceleration time profile see Appendix A
for 4 simple models, which are used throughout our analysis. For sake of simplicity,
we describe in this Section only Model 1.
Our simplest kinematic CME acceleration model is defined by,
\begin{equation}
        a(t) = \left\{ \begin{array}{ll}
        0       & \mbox{for $t < t_0$}         \\
        a_0     & \mbox{for $t_0 < t < t_A$}   \\
        0       & \mbox{for $t > t_A$}         \\
        \end{array}
        \right.
\end{equation}
with $\tau = t_A - t_0$.
From the acceleration profile $a(t)$ (parameterized with 2 variables $t_0$ and  
$a_0$) we can then directly calculate the velocity profile $v(t)$ by time
integration, which yields 
\begin{equation}
        v(t) = \int_{t_0}^t a(t) \ dt =
        \left\{ \begin{array}{ll}
        0                  & \mbox{for $t < t_0$}            \\
        v_0=a_0 (t-t_0)    & \mbox{for $t_0 < t < t_A = a_0 \tau $}      \\
        a_0 \tau           & \mbox{for $t > t_A$}            \\
        \end{array}
        \right.
\end{equation}
Equally we can derive the height-time profile $x(t)$ 
of the CME bulk mass by time integration of the velocity $v(t)$,
\begin{equation}
        x(t) = \int_{t_0}^t v(t) \ dt =
        \left\{ \begin{array}{ll}
        h_0                 & \mbox{for $t < t_0$}                      \\
        h_0+(1/2) a_0 (t-t_0)^2 & \mbox{for $t_0 < t < t_A$}            \\
        h_0+(1/2) a_0 (t-t_0)^2 + a_0\tau(t-t_A) & \mbox{for $t > t_A$} \\
        \end{array}
        \right.
\end{equation}
where we allow for an integration constant $h_0$ that represents the initial height
of the CME bulk mass. A good estimate of the initial height is the density scale height, 
$h_0 \approx \lambda = 50$ Mm $\times (T_e/1$ MK) as we determined from the preflare
temperature $T_e$ in Section 2.1.

The analytical expression of the velocity allows us also to calculate the
kinetic energy $E_{\mathrm{kin}}(t)$ of the bulk mass of a CME anytime on their trajectory,
using the mass $m_{\mathrm{cme}}$ (Eq.~13) and the CME velocity $v$ (Eq.~5),
\begin{equation}
        E_{\mathrm{kin}}(t) = {1 \over 2} m_{\mathrm{cme}}\ v^2(t) \ ,
\end{equation}
where the asymptotic velocity is $v_{\mathrm{cme}}(t \mapsto \infty)
=a_0 (t_A-t_0) = a_0 \tau$ for our standard Model 1 (Section 2.5), while the
corresponding values are $v_{\mathrm{cme}}=(1/2) a_0 \tau$ (Model 2), 
$v_{\mathrm{cme}}=(1/3) a_0 \tau$ (Model 3), and $v_{\mathrm{cme}}=a_0 \tau$ (Model 4), 
according to Eq.~(A3) in Appendix A. 

\subsection{The CME Radial Adiabatic Expansion Model }

The key assumption in our CME modeling is the concept of radial adiabatic expansion,
which provides a direct link between the height-time profile $x(t)$ (Eqs.~16, A4, A5) 
and the EUV
dimming profile $EM(t)$. We approximate the volume of the CME with a spherical shell
that initially is aligned with the stratified atmosphere (Fig.~3, top left), but
then progressively expands in radial direction (Fig.~3, top middle and right).
The radial volume expansion $V(t)$, starting from an initial volume $V_0$ with 
height $h_0$ can be parameterized as (Aschwanden 2009),
\begin{equation}
        {V(t) \over V_0} = {[R_{\odot} + h(t) ]^3 - R_{\odot}^3 \over
                            [R_{\odot} + h_0  ]^3 - R_{\odot}^3 }   \ .
\end{equation}
For adiabatic expansion, no energy or mass is exchanged between the interior and the exterior
of the (magnetically confined) CME volume, 
and thus conservation of the particle number dictates a reciprocal
relationship between the expanding volume and the decreasing mean electron density
inside the CME volume, i.e., $n_e \propto 1/V$, which predicts the following
evolution of the emission measure, 
\begin{equation}
		EM(t) \propto n_e(t)^2 V(t) \propto V(t)^{-1} \ ,
\end{equation}
which, combined with the volume parameterization given in Eq.~(18) yields
the following normalized emission measure ratio $q_{EM}(t)$,
\begin{equation}
        q_{EM}(t) = {EM(t) \over EM_0)} = {[R_{\odot} + h_0 ]^3   - R_{\odot}^3 \over
                              [R_{\odot} + h(t)]^3 - R_{\odot}^3 }   \ .
\end{equation}
The normalized emission measure $q_{\mathrm{EM}}(t)$ varies in the range of $[0,1]$ 
between the launch of the CME and its propagation to infinite distance ($h(t)=\infty$).
Including some constant emission measure $EM_{\mathrm{bg}}$ from a CME-unrelated background,
which we define as a fraction $q_{\mathrm{bg}}$ of the maximum emission measure 
$EM_{\mathrm{max}}=EM(t=t_{\mathrm{max}})$ during the flare/CME event, i.e., 
$EM_{\mathrm{bg}}=q_{\mathrm{bg}} EM_{\mathrm{max}}$, we generalize Eq.~(20) to,
\begin{equation}
	EM(t) = EM_{\mathrm{max}} [q_{\mathrm{bg}} + (1 - q_{\mathrm{bg}}) q_{\mathrm{EM}}(t) ] \ ,
\end{equation}
which varies from $EM(t_0) = EM_{\mathrm{max}}$ at the initial time $t=t_0$
of CME launch asymptotically to $EM(t=\infty)= q_{\mathrm{bg}}\ EM_{\mathrm{max}}$ after 
the CME event. 

Inserting the height-time profile $h(t)=x(t)$ (Eq.~16) into the emission measure profile
$EM(t)$ (Eq.~21), we have then a complete model (Fig.~4, bottom right) that can be fitted
to the observed dimming curve $EM_{\mathrm{obs}}(t)$ obtained from the DEM method. Since the
wavelength-dependent temperature effects are fully taken into account in the
DEM fits, the emission measure profile $EM(t)$ reflects directly the mass
dependence of the expanding CME and is not affected by any temperature effects,
such as radiative or conductive cooling. In other words, regardless whether the
plamsa inside the CME volume heats up or cools down, its number density is
fully accounted for in our temperature-integrated DEM in the temperature range of
$T_e \approx 0.5-20$ MK.

We can then forward-fit the model
based on the Eqs.~(16-21) and obtain the free variables $a_0$, $t_0$, and $q_{\mathrm{bg}}$,
as well as the kinetic energies $E_{\mathrm{kin}}$ of CMEs. Generally,
this EUV dimming model fits the data well.
Exceptions are large complex CME events with
multiple expansion phases, which require multiple convolutions of expansion
profiles. Other exceptions are the cases of ``failed CME'' events that 
produce insufficient energy to escape the solar corona. 

\subsection{	Gravitational Potential Energy of CMEs 	}

In addition to the kinetic energy we calculate the gravitational potential 
energy $E_{\mathrm{grav}}$ that is required to lift a CME from the solar
surface to infinity,
\begin{equation}
	E_{\mathrm{grav}}(h) = \int_{R_{\odot}}^\infty
	{G M_{\odot} m_{\mathrm{cme}} \over r^2 } dr 
	= { G M_{\odot} m_{\mathrm{cme}} \over R_{\odot}} 
	\ ,
\end{equation}
where $r$ is the distance of the CME (centroid) to the center of the Sun, 
$r_{\odot}$ is the solar radius, $M_{\odot}$ is the solar mass, and $G$ is the 
gravitational constant. The gravitational energy was found to be typically 
about an order of magnitude higher than the kinetic energy in the study of
Vourlidas et al.~(2000), and thus is an important quantity to include in estimates
of the kinetic energy of CMEs. 

It is also instructive to calculate the escape velocity, which we obtain from
setting the kinetic energy equal to the gravitational potential energy,
i.e., $E_{\mathrm{kin}} = (1/2) m_p v_{\mathrm{esc}}^2 = G M_{\odot} / R_{\odot}$, which
yields,
\begin{equation}
	v_{\mathrm{esc}} = \sqrt{ {2 G M_{\odot} \over R_{\odot} } }
	\approx 618\ {\rm km s}^{-1} \ .
\end{equation}
When applying the velocity model $v(t)$ (Eq.~15) to the observed CME
speeds $v_{\mathrm{cme}}$, we have to be aware that $a_0$ represents the
net acceleration rate, after subtracting the gravitational acceleration
$-g_{\mathrm{grav}}$,
\begin{equation}
	a_0 = a_{\mathrm{LF}} - g_{\mathrm{grav}} \ ,
\end{equation}
where $a_{\mathrm{LF}}$ represents the acceleration caused by the
magnetic Lorentz force. 

\subsection{	Numerical Code and Estimates of Uncertainties		}

After we described the theoretical aspects of our EUV dimming model
in the foregoing sections, we provide a brief summary of the numerical
code with estimates of uncertainties. The numerical code executes the
following tasks in sequential order:  

\begin{enumerate}
\item{\underbar{Data Acquisition:} (see Section 2.1 on data setup).
AIA/SDO images (Level 1.0) are read in all coronal wavelengths,
decompressed, and processed with AIA$\_$PREP to Level 1.5, which includes 
flat-fielding, removing of bad pixels, image coalignment between different 
wavelengths, rescaling to a common plate scale, and derotation (of the roll 
angle), so that the EW and NS axes are aligned with the horizontal and vertical
image axes $[x,y]$. The coalignment of level 1.5 data is believed to be
accurate to 0.5 pixels (SDO/AIA Data Analysis Guide). For our dimming
analysis we rebin the data to macro-pixels (with a size of $32\times 32$
full-resolution pixels), which yields $\approx 25^2=625$ macro-pixel 
locations for the spatial analysis of EUV dimming regions, for each of
the $n_t=10-200$ time sequences in the $n=399$ flare/CME events.}

\item{\underbar{DEM Analysis:} The automated spatial-synthesis DEM code 
(Aschwanden et al.~2013) performs in each of the 625 macro-pixel locations
a Gaussian forward-fit (with 3 parameters) to the 6 wavelength fluxes,
which is then synthesized to a combined DEM function $dEM(T,t)/dT$ for
each time step and event. Bad DEM solutions, which occur mostly due to
image saturation during the flare peak at too long exposure times,
are interpolated from the DEM solutions of the previous and following
time step. Bad DEM solutions are easily recognized from large 
$\chi^2$-values of the DEM goodness-of-fit criterion. The preflare temperature
is evaluated from the first image taken at the flare start time
$t_{\mathrm{start}}$ (defined by the NOAA flare catalog) and is always found
around $T_e \approx 1.5$ MK.}

\item{\underbar{EUV Dimming Time Profile:} We find that every EUV dimming
is preceeded by a previous increase in the total emission measure 
$EM_{\mathrm{tot}}(t)$, and the largest amount of dimming occurs always
at the location of the strongest previous emission measure increase.
In order to determine the location with the most significant dimming
thus requires the detection of the location with the maximum
emission measure in time and space, i.e., 
$EM_{\mathrm{tot,max}}(x_{\mathrm{max}}, y_{\mathrm{max}}, t_{\mathrm{peak}})$. 
In order to improve the statistics we calculate the cross-correlation coefficients 
between this maximum emission measure time profile and the remaining 724 
time profiles $EM_{\mathrm{tot}}(x,y,t)$, and combine all those emission
measure profiles that have a cross-correlation coefficient of
$CCC \ge 0.9$. The summed emission measure profile $EM_{\mathrm{tot,sum}}(t)$,
normalized to the total emission measure profile peak $EM_{\mathrm{tot}}(t_{\mathrm{peak}})$,
is then also corrected for temporal outliers by interpolation. The temporal
interpolation is found to be most effective in removing oscillatory fluctuations 
caused by the automated AIA exposure control, which is triggered by CCD saturation 
during flare peaks. The absolute uncertainty of the emission measure values 
in the time profile $EM_{\mathrm{tot}}(t)$, obtained with the spatial-synthesis DEM code,
is conservatively estimated to be of order $\approx 10\%$, based on a benchmark test with
simulated AIA data, which yielded an uncertainty in the peak emission measure of
$q_{\mathrm{EM,p}}=0.82\pm0.08$, and an uncertainty of the total (temperature-integrated)
emission measure of
$q_{\mathrm{EM,t}}=1.00\pm0.01$ (Aschwanden et al.~2013). Since we apply
multiple ($n_{\mathrm{model}}$) model fits, the resulting uncertainty is
$\sigma_q\approx 0.1/\sqrt{n_{\mathrm{model}}}$, which amounts to $\approx 5\%$
for out set of four models (Appendix A).}

\item{\underbar{Forward-Fitting to Dimming Curve:} The observed dimming curve
of the total emission measure $EM(t)$ includes data points from the rise time
of the flare, while EUV dimming generally starts after the flare peak. Therefore
we have carefully to define the time interval of fitting. Since the theoretical
dimming curve (Fig.~4 and Eqs.~20,21) predicts a monotonic decrease of the 
emission measure, we define the fitting time interval in the time range where
the total emission measure decreases from 90\% of the peak value down to 10\%
of the peak value (above the background level $q_{\mathrm{bg}}$). 
Then we fit the four dimming models (specified in Appendix A
and Fig.~4), which are characterized by the four free parameters $a_0, t_0, t_A,
q_{\mathrm{bg}}$, but keep the acceleration time interval $\tau = t_A - t_0$ fixed at 
the time resolution $\Delta t=2$ min, since we do not have a sufficient
number of datapoints that are sensitive to extract a variation in the
acceleration rate at this time resolution. The 3-parameter dimming models
(Eqs.~14-21) are fitted to the observed time profiles $EM_{\mathrm{obs}}(t)$ with
a Powell minimization method (Press et al.~1986, p.294). The best-fit model
based on least-square fitting (with estimated uncertainties of 
$\sigma_q\approx 0.1/\sqrt{n_{\mathrm{model}}}$ yields then the acceleration
profile $a(t)$, velocity time profile $v(t)$, and height-time profile $x(t)$.}

\item{\underbar{CME Parameters:} From the dimming profiles that show a decrease
of the EUV emission measure above some threshold level (defined by the two-fold
median value of emission measure differences before and after the dimming phase),  
we obtain the projected dimming area $A_p$ in the source region of the CME, which 
yields, together with the heliographic position $(l,b)$, the aspect angle $\rho$ 
(Eq.~9), and with the preflare background temperature $T_{\mathrm{bg}}$, the density 
scale height $\lambda$ (Eq.~3), and the unforeshortened length scale $L$ (Eq.~10).
From the total emission measure $EM_{\mathrm{tot}}$ at the flare peak time (i.e., of the
emission measure), we obtain then with the volume $V=L^2 \lambda$, the mean
electron density $n_e$ (Eq.~12) at the time of the CME onset, and the
mass of the CME (Eq.~13). The asymptotic CME velocity $v_{\mathrm{cme}}$ 
is then computed from the best-fit parameters $(a_0, \tau)$ (Eq.~15),
yielding also the kinetic energy $E_{\mathrm{kin}}$ (Eq.~17) for the CME.
The gravitational potential energy $E_{\mathrm{grav}}$ (Eq.~22) follows directly
from the CME mass $m_{\mathrm{cme}}$ (Eq.~22). }

\item{\underbar{Occurrence Frequency Distributions:} Power law functions
are fitted to the upper tails of the distributions, at $x > x_{\mathrm{max}}
=x(N_{\mathrm{max}})$, where the lower fitting boundary is chosen at the 
maximum of the size distribution, $N(x) \propto x^{-\alpha}$. The uncertainty
of the power law slope is estimated from $\sigma_\alpha = (\alpha - 1) / \sqrt{(n)}$,
with $n$ the number of events contained in the fitted part of the size
distribution, $n=\int_{x_{\mathrm{max}}}^{\infty} N(x) dx$ (Clauset et al.~2009).}

\item{\underbar{Comparison with LASCO:} Compiling our list of 399 EUV dimming
events with the existing LASCO CME catalog (Gopalswamy et al.~2009)
(http://cdaw.gsfc.nasa.gov/CME$\_$list), we compare the mass, speed, and
kinetic energy of CMEs for identical events.}

\end{enumerate}

\section{	OBSERVATIONS AND RESULTS 		}

We analyze observations from AIA/SDO for the 399 flare events of the
primary data set that consists of all solar GOES M and X-class flare
events observed with AIA/SDO during the first 3.5 years of the SDO mission 
(June 2010 - Jan 2014), which is identical to those analyzed in 
Papers I, II, and III for other forms of energies. The major CME parameters
obtained from our data analysis of the 399 events are compiled in
a {\bf machine-readable data file}, from which we reproduce only the first
10 entries in Table 3 and 4 of the printed publication, while the
complete data file can be downloaded from the electronic version of this
publication.

In Figs.~5-12 we
present the analysis of 24 exemplary events, which show the following
observations and results for each event: 
{\sl Top panels:} An AIA 193 \ang\ flux image
of the CME source region with a field-of-view of $FOV=0.5 R_{\odot}$,
observed at the peak time $t_{\mathrm{peak}}$ of the total emission measure profile
$EM_{\mathrm{tot}}(t=t_{\mathrm{peak}})$; {\sl Second row panels:} A difference
image between the AIA 193 \ang\ peak time and the following minimum of
the total emission measure profile $EM_{\mathrm{tot}}(t=t_{min})$,
with the contours of the dimming area $A$ overlaid as determined from
the spatial-synthesis DEM emission measure map; {\sl Third row panels:}
DEM distributions for each time step, marked with dashed curves before
the flare peak time $t_{\mathrm{peak}}$ and with solid curves after the flare peak;
{\sl Bottom row panels:} The total emission measure profile $EM_{\mathrm{tot}}(t)$
is indicated with a histogram with a time resolution of $\Delta t=2$ min,
the best-fit EM dimming curve with a red curve, fitted in the
10\%-90\% range of the emission measure variation (dotted vertical lines),
along with the flare start, peak, and end times according to the GOES flare 
catalog (vertical dashed lines).

\subsection{	Simple and Complex Events 		}

The emission measure dimming curves $EM_{\mathrm{tot}}(t)$ are shown for
12 ``simple events'' in Figs.~5-8, and for 12 ``complex events'' in
Figs.~9-12. What we mean with ``simple'' and ``complex'' events is
depicted in Fig.~13: A simple event consists of a short impulsive peak 
of the total emission measure, which is composed of a short rise time that
is immediately followed by a short decay time interval, closely matching 
the theoretically predicted EUV dimming behavior after
the emission measure peak time as displayed in Fig.~4 (bottom panel). 
A measure of the simple EUV dimming behavior is the dimming ratio,
\begin{equation}
	q_{\mathrm{dimm}}={(EM_{\mathrm{max}} - EM_{\mathrm{min}})
	            \over (EM_{\mathrm{max}} - EM_{\mathrm{bg}}) } \ ,
\end{equation}
which is close to unity for simple events. The selection of 12 events
displayed in Figs.~5-8 have all high dimming ratios of 
$q_{\mathrm{dimm}}\approx 0.92-0.96$ and acceptable goodness-of-fit values of
$\chi^2 \lapprox 1.2$, but otherwise exhibit a wide variety of parameters.
The typical CME speeds derived for simple events (see values labeled 
with $v$ in bottom panels of Figs.~5-8) occur in the range of 
$v_{\mathrm{cme}} \approx 200-1700$ km s$^{-1}$, which
is consistent with the typical CME velocity range reported from 
white-light observations. These examples of ``simple events''
show a goodness-of-fit with reduced chi-square fits of 
$\chi\lapprox 1.2$, and thus demonstrate that the radial adiabatic 
expansion model (Section 2.6) yields reasonable CME speeds and
acceptable fits. The spatial scales
of simple events shown in Figs.~5-8 vary in the range of
$L \approx 80-250$ Mm, and the associated flare durations show
a range of $D \approx 0.1-0.5$ hrs. The variety of simple events 
shown in Figs.~5-8 includes events observed near disk center 
(Fig.~5 left and right; Fig. 8 left and middle), near the limb 
(Fig.~5 middle, Fig.~6 right, Fig.~8 right), events with CCD saturation
at the flare peak time (Fig.~7), and thus demonstrates that simple events
do not occur at particular heliographic positions, and are retrieved
even in the case of image saturation. In summary,
the simple events reveal the following characteristics of EUV dimming:
(i) the emission measure in the EUV dimming region increases by a 
large amount during the soft X-ray flare time interval, (ii) and
drops by a large amount (of $\gapprox 0.90\%$) after the flare,
and (iii) the dimming profile closely follows the theoretically
predicted function that is expected for radial adiabatic expansion.
In fact, we found a highly significant level of EUV dimming in all
of the 399 analyzed flares, so there is no sensitivity issue in 
measuring the EUV dimming in M and X-class flares.

Let us go to the other extreme of ``complex events'', which we
interpret as events that consist of multiple peaks that are
convolved with each other (Fig.~13 bottom panel) and may hamper
the deconvolution of simple EUV dimming behavior. In Figs.~9-12
we present a selection of 12 events that were selected in order
of the longest flare durations ($D \gapprox 1.0$ hr) and having 
acceptable fits ($\chi^2 \lapprox 1.2$).
Long-duration flares are inherently more complex in the structure
of their light curves, showing often multiple peaks that cannot
easily be deconvolved from each other. Our simple model of a single
dimming phase may not be adequate for such complex events, but we
fitted the dimming curve by the same rules, in a fitting time interval
that encompasses the 10\%-90\% range of the EUV dimming.
In some sense, these examples shown in Figs.~9-12 represent the worst 
cases, while the examples in Figs.~5-8 represent the most 
ideal cases for fitting of our model. In the complex events we
notice step-wise jumps in the emission measure,
which are likely to be caused by CCD saturation and the automatic 
exposure control mode, data drop outs during CCD saturation,
multiple peaks, or gradual (rather
than impulsive) transitions of emission measure increases to
decreases (Figs.~9-12). We notice also that the dimming
decreases are significantly lower for complex events than for
simple events, in the range of $q_{\mathrm{dimm}} \approx 45\%-85\%$.
Moreover, the resulting CME velocities
are significantly lower for complex events than for simple events,
in the range of $v_{\mathrm{cme}} \approx 30-120$ km s$^{-1}$ for the 12 cases
shown in Figs.~9-12, which are partially caused by superimpositions
of multiple dimming phases that cannot easily be deconvolved
in the EUV dimming profiles.

The fact that the fitted CME speeds are systematically lower for
complex events of longer duration than for simple events,
may indicate a convolution bias of multiple dimming phases that
are triggered at multiple times during the soft X-ray flare.
The observed EUV emission measure time profile $EM(t)$ in a complex
event with duration $D$ can be considered as a convolution of 
$N$ elementary time profiles with duration $d$ in simple events
(Fig.~13), which are related to each other by Poisson statistics for a
random process,
\begin{equation}
	D = \sqrt{N} d \ .
\end{equation}
The CME speed $v$ is proportional to the slope of the emission measure 
decrease $dEM/dt$, since,
\begin{equation}
	{dEM(t) \over dt} = {dEM(t) \over dh} {dh \over dt}
	\approx {\Delta EM \over \Delta h} \ v  \ .
\end{equation}
Low velocities determined with the EUV dimming method may therefore
contain a convolution bias of multiple (spatially or temporally
separated) dimming regions. Quantitative analysis of low-speed CME events 
may require more complex modeling with multiple dimming components,
which is beyond the scope of this study.  

\subsection{	Correlations of Geometric CME Parameters 	}

We investigate statistical correlations of CME parameters that we measured
from the 399 AIA events, which are presented in form of scatter plots
in Fig.~14, with the parameter ranges and distributions compiled in Table 1.
We start with the geometrical parameters. In our ``square-box''
model of the CME source region (Eq.~5) we find a strong correlation 
between the observed projected area $A_p$ and the inferred deprojected
CME source area $A$, with $A \propto A_p^{1.5\pm0.1}$ 
and a regression coefficient
of $R=0.93$ (Fig.~14a), which is a statistical relationship that
averages over all heliographic positions, while the exact relationship
is given by Eqs.~(5-10). The projection effect yields also 
$A_p \propto L^{1.3\pm0.1}$ (Fig.~14b), in contrast to the approximation
$A_p \approx A \propto L^2$ that is expected when neglecting projection
effects. Also the volume scales as $V \propto L^{1.98\pm0.02}$ (Fig.~14c), due
to the constant vertical scale height in gravitationally stratified
atmospheres, rather than the often used approximation $V \approx L^3$.
In order to correct for projection or foreshortening effects one
can use either these empirical scaling relationships, or the exact
relationships given in Eq.~(5-10) if the heliographic position of the
dimming region is known.

Since the emission measure $EM_{\mathrm{tot}} \propto n_e^2 V$ scales with the
volume $V$, we expect some correlation, which is indeed confirmed by
the data (Fig.~14d). A correlation of $EM_{\mathrm{tot}} \propto V$ is 
expected since the preflare electron density has a very
narrow distribution, i.e., $n_e=(1.3 \pm 0.6) \times 10^9$ cm$^{-3}$
and does not show any correlation with the CME source volume (Fig.~14e).
Similarly, the preflare electron temperature is found to have a 
very small variation also, i.e., $T_e=(1.8 \pm 0.3)$ MK (Fig.~14f), which
is not correlated with any spatial scale.

Owing to the small variation in electron density, the CME mass 
$m_{\mathrm{cme}}$ is expected to be highly correlated with the CME source
volume $V$ (Eq.~13), which indeed is the case, yielding 
$m_{\mathrm{cme}} \propto V^{1.07\pm0.06}$ with a regression coefficient of
$R=0.95$ (Fig.~14g). Similarly, the CME kinetic energy (Eq.~17)
shows the expected correlation $E_{\mathrm{kin}} \propto v^2$
(Fig.~14h). However, neither the CME speed nor the CME kinetic energy
exhibits any correlation with a spatial scale (Fig.~14j and 14k), 
which indicates that the acceleration and speed occurs in vertical direction
and are not physically coupled to the horizontal extent $L$ of the CME,
due to magnetic confinement. Also the dimming fraction is not correlated with
the horizontal extent of the dimming region (Fig.~14i), probably 
because the CME expansion occurs initially only in vertical direction
due to the magnetic confinement in the low plasma-$\beta$ regions
of the corona.

\subsection{	Correlations of Temporal CME Parameters  		}

We measure two temporal parameters of CMEs, namely the half-dimming
time $\tau_{\mathrm{dimm}}$ and the vertical CME propagation time
$\tau_{\mathrm{prop}}$. We define
the half-dimming time interval $\tau_{\mathrm{dimm}}$ as the interval between 
the dimming onset time $t_0$ (from our fits of the dimming models) and the 
time when the emission measure drops to 50\% of the (normalized) 
dimming fraction $q_{\mathrm{EM}}(t)$ (Eq.~20). The vertical CME propagation
time scale $\tau_{\mathrm{prop}}=\lambda/v$ is simply the vertical density
scale height divided by the CME speed $v$. Both time scales have
similar values in our sample of 399 events, i.e., 
$\tau_{\mathrm{dimm}}=0.5-33$ min, versus
$\tau_{\mathrm{prop}}=0.3-54$ min (Table 1), and are highly correlated, 
with a scaling of $\tau_{\mathrm{prop}} \propto \tau_{\mathrm{dimm}}^{1.5\pm0.3}$,
and a linear regression coefficient of $R=0.80$ (Fig.~15a).
However, none of these CME time scales is correlated to the
horizontal length scale of the dimming region (Fig.~15b, 15c),
which again suggests dominant vertical transport of the CME
mass without horizontal diffusion.

When we consider thermal time scales, such as the GOES rise time,
decay time, or flare duration, which all are highly correlated
among each other, we find that they are
also correlated with the CME time scales. The best correlation
among the 3 GOES time scales is found for the GOES decay time
with the AIA half-dimming time 
($\tau_{\mathrm{dimm}} \propto \tau_{\mathrm{decay}}^{1.0\pm0.3}$, with a regression
coefficient of $R=0.69$; Fig.~15f), and the AIA CME
propagation time
($\tau_{\mathrm{dimm}} \propto \tau_{\mathrm{decay}}^{1.4\pm0.4}$, with a regression
coefficient of $R=0.72$; Fig.~15i). This is a remarkable result,
because the correlated parameters are measured with different
instruments (GOES and AIA), in different wavelength domains,
and from different physical processes.
The GOES decay time is supposedly determined by plasma heating and cooling
time scales, while the AIA half-dimming time scale and
propagation time scale is governed by plasma transport during 
an adiabatic expansion process. Our result may indicate that adiabatic
expansion may also play an important role in the interpretation
of flare decay times measured in soft X-rays (with GOES), 
besides the processes of plasma cooling by thermal conduction 
and radiative loss.

How is the CME speed $v$ related to temporal parameters? The scatter
plots shown in Fig.~15j, 15k, and 15l clearly show that the (final)
CME speed is approximately reciprocally related to all time scales,
the GOES flare duration, the AIA half-dimming time, and the AIA CME 
propagation time. This is expected for the AIA CME propagation time
due to its definition, i.e., $\tau_{\mathrm{prop}}=h_0/v$, and as a consequence
of the correlation of the CME dimming time (Fig.~15d) or the CME
propagation time (Fig.~15g) with the GOES flare duration. Partially,
these correlations are related to the convolution bias of ``complex
events'', as depicted in Fig.~13: The longer the flare duration,
the longer the convolved gradient $dEM/dt$ of the dimming profile,
which implies a lower velocity $v$ and a longer dimming half time 
$\tau_{\mathrm{dimm}}$. 

\subsection{	Occurrence Frequency Distributions 		}

The occurrence frequency distributions of the CME parameters
measured with AIA are shown in Fig.~16 and compiled in Table 2. 
These size distributions
show power law-like functions for most CME parameters, such as
the length $L$ (Fig.~16a), the CME source volume $V$ (Fig.~16b),
the flare duration $D$ (Fig.~16c), the total emission measure $EM_{\mathrm{tot}}$
(Fig.~16d), the CME speed $v$ (Fig.~16e), the CME mass $m$ (Fig.~16f),
the CME kinetic energy $E_{\mathrm{kin}}$ (Fig.~16g), the CME gravitational
potential energy $E_{\mathrm{grav}}$ (Fig.~16h), and the total kinetic plus
gravitational energy $E_{\mathrm{cme}}$ (Fig.~16i). In contrast, the electron
density $n_e$, the preflare electron temperature $T_e$, and
the dimming fractions $q_{\mathrm{dimm}}$ (Fig.~16j) have narrow distributions
that steeply fall off without power law tail.

The slopes of the power laws can be predicted from self-organized
criticality (SOC) models, with some model modification for CME phenomena.
The most fundamental prediction of SOC models is the {\sl scale-free
probability conjecture}, which states that the statistical probability
of the occurrence rate $N(L)$ of nonlinear energy dissipation avalanches
scales reciprocally to the geometric scale $L$ of the phenomena 
(Aschwanden 2012),
\begin{equation}
	N(L) \propto L^{-d} \ ,
\end{equation}
where $L$ is the geometric length scale and $d$ is the Euclidean
space dimension, which for most real-world phenomena is generally $d=3$.
The distribution of CME length scales, i.e., 
$N(L) \propto L^{-(3.4 \pm 1.0)}$ (Fig.~16a), appears to be consistent
with the scale-free probability conjecture, $N(L) \propto L^{-d}$, 
within the uncertainties of the power law fit.

From the length distribution $N(L) \propto L^{-3.4}$ and the theoretical
CME volume scaling $V \propto L^2 \lambda \approx L^2$ (Eq.~5, Fig.~14c)
we can directly predict the volume size distribution $N(V)$, using
$L[V] \propto V^{1/2}$ and $dL/dV=V^{-1/2}$,
\begin{equation}
	N(V) \propto N(L[V]) {dL \over dV} \propto V^{-2.2} \ ,
\end{equation}
which is indeed consistent with the measured size distribution
$N(V) \propto V^{-2.2\pm0.5}$ (Fig.~16b). Note the modification of
standard models $V \propto L^3$, to $V \propto L^2$ for CME source
volumes in a gravitationally stratified atmosphere, which is a
consequence of similar scale heights $\lambda$ for all CMEs. 

Using the approximation of $m \propto n_e V \propto V$ for CME masses
(due to the small variation of mean electron densities $n_e$) we expect
that the size distribution of CME masses is identical to those of the
volumes, which is indeed the case, i.e., $N(m) \propto m^{-2.2\pm0.4}$
(Fig.~16f). In the same vain we expect the same distribution for
total emission measures, since $EM \propto n_e^2 V \propto V$ (Eq.~11), 
which is indeed the case, i.e., $N(EM) \propto EM^{-2.4\pm0.4}$
(Fig.~16d).

It is not obvious how the length scale $L$ is related to the event
duration $D$ for CMEs, since there is no significant correlation
(Fig.~15b). However, a theoretical prediction of the standard SOC model 
is that the spatio-temporal relationship is approximatedly a 
diffusive random walk that obeys the diffusion equation (Aschwanden 2012),
\begin{equation}
	L = \kappa D^{1/2} \propto D^{1/2} \ ,
\end{equation}
where $\kappa$ is the diffusion coefficient, assuming to be uncorrelated
with the event duration. 
Combining the two fundamental assumptions (Eq.~28, 30) predicts
then the following size distribution for CME event durations $D$,
using $L \propto D^{1/2}$ and $dL/dD \propto D^{-1/2}$,
\begin{equation}  
	N(D) \propto N(L[D]) {dL \over dD} \propto D^{-2.2} \ ,
\end{equation}
which indeed is consistent with the observed distribution,
$N(D) \propto D^{-2.5\pm0.6}$ (Fig.~16c). We use here the dimming
time as a measure of the event duration, $D \propto \tau_{\mathrm{dimm}}$,
in order to compare EUV observables self-consistently, but the same
result is also obtained by using GOES flare durations $D$.

Another fundamental relationship we found in our study is the
relationship between the CME speed $v$ and the CME event
duration duration $D$, which scales with $v \propto D^{-1.6\pm0.6}$
(Fig.~15j). Combining this relationship with the distribution
of event durations, we predict,
\begin{equation}  
	N(v) \propto N(D[v]) {dv \over dD} \propto v^{-1.8} \ ,
\end{equation}
which is indeed consistent with the observed size distribution
of CME speeds, i.e., $N(v) \propto v^{-1.9\pm0.3}$ (Fig.~16e). From this we
can also estimate the size distribution of CME kinetic energies,
using $E_{\mathrm{kin}} \propto v^2$, $dv/dE_{\mathrm{kin}} \propto E_{\mathrm{kin}}^{-1/2}$,
\begin{equation}  
	N(E_{\mathrm{kin}}) \propto N(v[E_{\mathrm{kin}}]) 
	{dv \over dE_{\mathrm{kin}}} \propto v^{-1.4} \ ,
\end{equation}
which matches the observed size distribution, i.e.,
$N(E_{\mathrm{kin}}) \propto E_{\mathrm{kin}}^{-1.4\pm0.1}$ (Fig.~16g). 

For the gravitational energy $E_{\mathrm{grav}}$, we expect a size 
distribution that is nearly identical to that of the masses,
since $E_{\mathrm{grav}} \propto m$, which is indeed the case,
i.e., $N(E_{\mathrm{grav}}) \propto E_{\mathrm{grav}}^{-(2.2 \pm 0.5)}$ 
(Fig.~16h), versus $N(m) \propto m^{-2.2\pm0.5}$ (Fig.~16f).

Finally, for the total CME energy, which is defined by the sum
$E_{\mathrm{tot}}=E_{\mathrm{kin}}+E_{\mathrm{grav}}$, we expect an approximate slope
that is close to the average of the two power law slopes,
i.e., $a_{\mathrm{tot}} \approx (a_{\mathrm{kin}} + a_{\mathrm{grav}})/2 \approx 
(1.6 + 2.2)/2 = 1.9\pm0.3$, which matches the observed size
distribution also, i.e., $N(E_{\mathrm{tot}}) \propto E_{\mathrm{tot}}^{-2.0\pm0.3}$
(Fig.~16i).

In summary, most of the observed size distributions of CME
parameters exhibit a power law-like function, and are consistent
with the scale-free probability conjecture and the random-walk
assumption of a fractal-diffusive avalanche model of a
slowly-driven self-organized criticality system
(Aschwanden 2012). However, two modifications are needed for
CME phenomena, namely the CME source volume scaling 
$V \propto L^2$ in a gravitationally stratified atmosphere, 
and the convolution bias of adiabatic expansion velocities, 
$v \propto D^{-1}$, which scales reciprocally with the
duration of flare/CME events. The predicted and observed
power exponents of the size distributions of CME parameters
and of the underlying parameter correlations are juxtaposed 
in Table 2. 

\subsection{	Comparison with LASCO Observations 		}

While we described the quantitative CME results based on the EUV dimming
analysis made with AIA data so far, we turn now to comparisons with
white-light results from identical CME events observed with LASCO 
onboard SOHO. A LASCO CME catalog was generated that contains 
the mass, speed and kinetic energy of CMEs, which is publicly 
available at http://cdaw.gsfc.nasa.gov/CME$\_$list, covering almost 
two decades of continuous LASCO observations, 1996 Jan-2015 July
(Gopalswamy et al.~2009). 

The first task is to identify identical events between the GOES
flare list (used for identifying EUV dimming events with AIA) and
the LASCO CME list, which is based on the delayed detection after
the CME emerges from behind the C2/coronagraph occulter disk at a
distance of $2.2 R_{\odot}$ solar radii. Sometimes, LASCO events
are detected with C3 only, which causes additional delays. 
The time delay between a CME-associated flare and the CME detection
with LASCO thus depends on both the heliographic location $(l,b)$ 
on the solar disk and the distance of first detection with LASCO
(after passing the occulter disk at $>2.2 R_{\odot}$). Denoting
the GOES flare peak time with $t_{\mathrm{peak}}$, the time of first 
detection with LASCO with $t_{\mathrm{det,C2}}$, and the distance from
Sun center at the time of first detection with $x_{\mathrm{det,C2}}$, the
detection delay is defined as,
\begin{equation}
	\Delta t_{\mathrm{det,C2}} = t_{\mathrm{det,C2}} - t_{\mathrm{peak}} 
		= {(x_{\mathrm{det,C2}} - x_{\mathrm{flare}}) \over v_{\mathrm{LE}}} \ ,
\end{equation}
where $v_{\mathrm{LE}}$ is the average speed of the CME leading edge,
and $x_{\mathrm{flare}}$ is the distance of the flare site from Sun disk
center. Implicitly we assume that the propagation of the CME
leading edge is isotropic in the high plasma-$\beta$ region
of the upper corona and heliosphere.
For the speed $v_{\mathrm{LE}}$ of the leading edge we use the linear fit
to the height-time measurements at the first time of LASCO/C2
detection as listed in the LASCO CME catalog.
The travel distance to first detection is 
nominally $x_{\mathrm{det,C2}}=2.2 R_{\odot}$ for LASCO/C2, but can be larger
when a data gap occurred before CME detection, or shorter when
the flare occurred near the limb. The distance
of the flare site from Sun center depends on the heliographic
longitude $(l)$ and latitude $(b)$ of the flare site and is given
by the trigonometric relationship, 
\begin{equation}
	\cos{(\rho)} = cos{(l)} \cos{(b)} \ ,
\end{equation} 
where $\rho$ is the heliographic angle between the flare site
and Sun center, which translates into a distance by,
\begin{equation}
	x_{\mathrm{flare}} = R_{\odot} \sin{ (\rho) } \ .
\end{equation}

The distribution of propagation distances $(x_{\mathrm{det,C2}} - x_{\mathrm{flare}})$
is shown in Fig.~17a, ranging from 1.2 to $\gapprox 4.0$ solar
radii. The distribution of LASCO detection delays 
$\Delta t_{\mathrm{det,C2}} = (t_{\mathrm{det,C2}} - t_{\mathrm{start}})$ is shown in
Fig~17b, which varies in a range of $\approx 10-100$ min and
has a median value of 48 min. The ambiguity of identical events
detected with GOES and LASCO depends on the accuracy of LASCO
velocity measurements, required to extrapolate the timing.

In Fig.~17 we show the distributions of various time delays.
The time difference between the LASCO-extrapolated CME onset time
$t_{\mathrm{onset}}$ and the GOES flare peak time $t_{\mathrm{peak}}$ shows a
relatively broad distribution with a mean and standard deviation of
$\Delta t= -2 \pm 46$ min (Fig.~17c). We are not sure how many of 
these LASCO events within this temporal coincidence margin are
associated with the GOES and AIA-detected CME dimming events, but
the fact that the AIA-inferred dimming time coincides with the
GOES flare peak time within a much smaller time margin of 
$\Delta t=5 \pm 7$ min (Fig.~17e) indicates that many of the LASCO
events with delays $\gapprox 10$ min may not be associated with
the GOES and AIA-selected events. 

A comparison of the occurrence frequency distributions of AIA and
LASCO-inferred CME events is shown in Fig.~18 (left panels), for
the CME mass (top left panel), the CME speed (middle left panel),
and the CME kinetic energy (bottom left panel), and the
corresponding scatter plots between AIA and LASCO-inferred
quantities are shown in the right-hand panels of Fig.~18.
There is a overall agreement in the median values, but the
corresponding quantities measured with both instrument scatter
by up to two orders of magnitude. This lets us to conclude that
the the parameters are differently defined for the two instruments.
The method of measuring CME speeds with LASCO is designed to detect
the leading edge of a CME, because this is the location where the
largest density contrast occurs in white light, while the method
of measuring speeds with AIA is weighted by the bulk plasma,
which contributes most to the EUV emission measure. Since the
leading edges are faster than the bulk speed, we suspect that
LASCO obtains faster CME speeds than AIA, and consequently 
yields larger kinetic energies also. On the other side, LASCO
detects a number of CMEs with much lower masses than AIA, in
the range of $m_{\mathrm{cme}} \approx 10^{13}-10^{15}$ g, which appear
to be underestimates due to sensitivity issues in detecting
white-light polarized brigthness signals. We would expect that
LASCO detections underestimate the mass and velocities of halo
CMEs, but we are not able to find a center-to-limb effect in
the LASCO data.  

\subsection{	Comparison with GOES Data			}

Since the GOES fluxes are often used to characterize the magnitude
of solar flares, we investigate the question whether they are also
suitable to express the magnitude of CMEs. 
Scatter plots of CME masses, speeds, and kinetic energies are
compared with GOES 1-8 \ang\ fluxes for both LASCO and AIA 
observations in Fig.~19. We find that the best correlation
exists between the CME mass (as determined with AIA) and the
GOES flux, with an almost linear relationship, i.e.,
$ m \propto F_{\mathrm{GOES}}^{0.8}$, with a linear regression coefficient
of $R=0.67$ (Fig.~19b). The CME masses determined with LASCO reveal a
similar correlation, i.e., $m \propto F_{\mathrm{GOES}}^{0.7}$ (Fig.~19a), but
with a less significant linear regression coefficient of
$R=0.32$, indicating more random scatter in the uncertainties
of the CME masses. A similar correlation between CME masses
and GOES fluxes was found by Aarnio et al.~2011).

There is also a weak correlation between
the CME kinetic energies and the GOES fluxes (Figs.~19e and
19f), which is a consequence of the mass dependence of kinetic
energies. Nevertheless, these correlations reveal that a proportional 
amount of (magnetic) energy goes into plasma heating (as measured
in the soft X-ray plasma with GOES) and into the kinetic energy
of CMEs (as measured with AIA and LASCO). 

\subsection{	CME Energetics from EUV Dimming 		} 

Finally we compare the energy of CMEs with other forms of energies
in the overall flare energy budget, such as with the dissipated
magnetic energy $E_{\mathrm{mag}}$ in flares (Paper I and Figs.~20a and 20b), 
the free energy $E_{\mathrm{free}}$ (Paper I and Figs.~20d and 20e), 
the multi-thermal energy $E_{\mathrm{th}}$ (Paper II and Figs.~20g and 20h), 
and the nonthermal energy $E_{\mathrm{nth}}$ (Paper III and Figs.~20j and 20k).
We compare the CME kinetic energy from LASCO data (Figs.~20a, 20d,
20g, 20j), the CME kinetic energy from AIA data (Figs.~20b, 20e,
20h, 20k), and the total (combined kinetic and gravitational)
energy as measured from AIA data (Figs.~20c, 20f, 20i, 20l).
From this comparison shown in Fig.~20 we see that the CME kinetic
energy, regardless whether it is determined with LASCO or AIA,
exhibits a large scatter with any form of other energies, while
the total energy that includes both the kinetic energy and the
gravitational potential energy of a CME combined reveals a
much smaller scatter (down to a factor of three in the energy 
ratio). Therefore, it is imperial to include both the kinetic
and gravitational energy in comparisons with other flare-generated
energies, which is shown in the right column of Fig.~20.

According to these results we find that the total CME energy has
a logarithmic mean ratio of $E_{\mathrm{cme}}/E_{\mathrm{mag}}=0.07$ (with a scatter by a 
factor of 5.4; Fig.~20c), but is virtually always lower that the dissipated
magnetic energy in flares, as estimated from nonlinear force-free
field modeling (Paper I). A similar ratio is found for the free
magnetic energy, namely $E_{\mathrm{cme}}/E_{\mathrm{free}}=0.11$ (with a scatter 
by a factor of 6.3; Fig.~20f), which is a consequence of the fact
that the dissipated magnetic energy was found to be a large fraction
of the available free energy, at least in GOES M and X-class flares
(Paper I). The most interesting result is the ratio of the total
CME energy with the multi-thermal flare energy (Paper II), for which 
we find the tightest correlation, with a mean logarithmi ratio of 
$E_{\mathrm{cme}}/E_{\mathrm{th}}=0.77$ (with a scatter by a only a factor of 3.5;
Fig.~20i). This indicates that the energy that goes into a CME
is most closely related to the heating of the flare plasma.
This correlation is also consistent with the other correlation
found between the CME mass and the GOES flux (Fig.~19b).
A similar ratio is found for the ratio of the nonthermal
energy, namely $E_{\mathrm{cme}}/E_{\mathrm{nth}}=0.72$ (with a scatter by a 
factor of 9.8; Fig.~20l), calculated from the thick-target model
of accelerated and precipitating electrons with a fixed low-energy
cutoff of $20$ keV. 

\section{               DISCUSSION 				}

\subsection{    Previous EUV Dimming Observations               }

The EUV dimming method developed here is an alternative method to the 
CME mass determination by Thomson scattering in white-light observations.
Essentially, a reduction of the EUV brightness (or emission measure)
anywhere on the solar disk
or above the limb can be converted into a corresponding coronal mass loss,
carried away by a CME, by an eruptive filament, or by drainage down to the
chromosphere (coined ``coronal rain''; Schrijver 2001).
Early pioneering observations of coronal dimmings during energetic CMEs,
using SOLWIND and LASCO/SOHO, are described, e.g., in
Jackson and Hildner (1978), Howard et al.~(1985), and Thompson et al.~(2000).
A statistical study found that 55\% of EUV dimming
events detected with SMM/CDS are associated with CMEs detected in
LASCO data (Bewsher et al.~2008). {\it Vice versa}, 84\% of
CME events could be tracked to EUV dimming regions 
(Bewsher et al.~2008). EUV dimmings occur in 74\% of flares associated with CMEs
(Nitta et al.~2014).
In contrast, so-called ``stealth-CMEs'' have been
reported in a few cases (Robbrecht et al.~2009; Howard and Harrison
2013; Nieves-Chinchilla et al. 2013), which apparently leave the
solar corona ``with no trace left behind''.
In a study of 96 CME-associated EUV coronal dimming events using LASCO,
the time profile of the EUV dimming was characterized with two phases,
an initial rise and decay with similar time scales, followed by a flatter
decay lasting several hours (Reinard and Biesecker 2008), which we
model with a unified adiabatic expansion model in this study.

With the advent of STEREO, the 3D structure of CME source regions
and associated EUV dimming could be observed with the EUVI/A+B imagers
and modeled with stereoscopic methods (Aschwanden et al.~2009a,b;
Aschwanden 2009; Temmer et al.~2009; Bein et al.~2013).
A key result was that the CME mass determined from EUV dimming agreed
well with those determined with the white-light scattering method
($m_{\mathrm{EUVI}}/m_{\mathrm{LASCO}}=1.1 \pm 0.3$), and agreed also between the two
STEREO spacecraft A and B ($m_A/m_B=1.3 \pm 0.6$) (Aschwanden
et al.~2009a). Another benefit of stereoscopic observations is the
determination of the 3D trajectory and de-projected CME speed and mass
(Bein et al.~2013).

The most recent observations of EUV dimming use SDO/AIA images in
multiple wavelengths, which allows for differential emission measure (DEM)
analysis of the density and temperature structure of CMEs (Cheng
et al.~2012; Mason et al. 2014). From such studies it was concluded
that the EUV dimmings are largely caused by the plasma rarefaction
associated with the eruption (Cheng et al.~2012), which we model
in terms of adiabatic expansion in this study. Automated searches
of flares, EUV dimmings, and EUV waves in SDO/AIA images are now
available also (Kraaikamp and Verbeeck 2015).

\subsection{	White-light versus EUV Dimming Detection Efficiency   } 

In this study we developed a novel EUV dimming method that is capable
to measure the timing, mass, acceleration, speed, propagation, and
the kinetic energy of CMEs. The event selection in our statistical
study encompasses all ($N=399$) GOES M and X-class flares during the 
initial 3.5-year period of AIA/SDO and we found that the EUV dimming 
method yielded a positive detection and measurements of CME parameters 
in 100\% of the analyzed events. For the simultaneous detection of 
events with the white-light method using LASCO/C2, we find an upper
limit of possible detections for 69\% (with a time delay of
$\Delta t = (t_{\mathrm{LASCO}}-t_{\mathrm{start}}) = -2 \pm 46$ min between the
LASCO-extrapolated starting time at the coronal base and the 
GOES flare peak time; Fig.~17c), and a lower limit of 30\% for a delay 
range of $\Delta t = -0.3 \pm 7.3 $ min that is equivalent to the delays of
the AIA detections, with $\Delta t = t_{\mathrm{dimm}}-t_{\mathrm{start}}=5 \pm 7$ min, 
Fig.~17e) between the onset of EUV dimming and the GOES flare peak time. 
Therefore, the EUV dimming method is found to be far more sensitive
to detections of CMEs, and we can expect positive detections for
weaker flares (of GOES C-class or lower). We know that there exist
many CMEs that are associated with weaker flares than GOES M1.0 class,
because the LASCO CME catalog shows many entries in the time intervals
between the M and X-class events analyzed here. This is also consistent
with the statistics of an EUVI/STEREO study, which included 185 flare
events and where 31 EUV dimming events were identified, mostly C-class
events, while only 11 events were of M-class (Aschwanden et al.~2009b).

On the other side, however,
there are a few cases of ``stealth CMEs'' known, which apparently
were detected in white-light but not by EUV dimming (Robbrecht 
et al.~2009; Howard and Harrison 2013; Nieves-Chinchilla et al.~2013).
It is conceivable that the larger temperature range of AIA 
($T_e \approx 0.5-20$ MK) could reveal EUV dimmings of stealth CMEs that
are not detectable in the narrower temperature ranges of EIT/SOHO or 
EUVI/STEREO (Robbrecht et al.~2009; Howard and Harrison 2013),
or that CME-associated flare sites were occulted behind the limb. 

\subsection{	The Measurement of CME Speeds 			  } 

Since we find a large scatter between CME speeds determined with the
white-light method and with the EUV dimming method (Fig.~18) we need
to understand whether the differences result from different physical
speed definitions or from uncertainties in the measurement methods.

CMEs observed in white-light are detected from difference images in
order to subtract out the unpolarized static white-light scattering 
component (of the entire Sun) from the polarized brightness produced
by the CME. The features that are brightest in difference images are
always where the steepest brightness gradients occur, which is
generally at the leading edge of expanding CMEs, and therefore
the inferred CME speed is actually the leading-edge speed $v_{\mathrm{LE}}$,
which is supposedly the fastest speed by definition. White-light
observations of CMEs often show a ``three-part'' structure, consisting
of (1) a bright front, (2) a dark cavity, and (3) a bright, compact
core. The bright front corresponds to the leading edge. The position
of the leading edge is then measured in terms of a height-time plot
$h(t)$, while the speed $v(t)=dh(s)/dt$ is derived from various
fitting functions. The speeds that are compiled in the LASCO CME
catalog were derived with three different functions: a linear or
first-order polynomial fit $v_1$, a quadratic or second-order fit 
$v_2$, and a speed $v_3$ evaluated at a distance of 20 solar radii. 
Comparing the 228 LASCO events with triple-velocity measurements 
at the earliest time of LASCO/C2 detection, out of the 399 GOES flare 
events analyzed here, we find ratios of $v_1/v_2=1.0\pm0.2$,
so we assess an uncertainty of $\approx 20\%$ to the white-light
CME leading edge speed measurements. We have also to be aware that
white-light speeds are projected speeds $v_{\mathrm{proj}}$ in the 
plane-of-sky, which should be corrected by 
$v = v_{\mathrm{proj}}/ \sin{( l )}$ at a heliographic longitude $l$,
as it was done for some stereoscopic measurements to derive the
true CME speed and mass (Colaninno and Vourlidas 2009). 

The speed measurements with AIA are conducted with a completely
different method, based on the best fits of EUV dimming light
curves $EM_{\mathrm{tot}}(t)$ (Eq.~20), which yields a best-fit height-time
profile $x(t)$, from which a speed $v = dx(t)/dt$ is derived.
Thus, the gradient in the dimming curve $dEM_{\mathrm{tot}}/dt \propto v(t)$
is decisive for the speed measurements. Obviously, the largest
contributions to the EUV dimming curve come from the locations
with the largest emission measures, and thus it represents a
bulk plasma speed $v_{\mathrm{bulk}}$. Since the leading edge is by
definition faster than the bulk plasma speed, $v_{\mathrm{LE}} > v_{\mathrm{BP}}$,
we expect that the LASCO-inferred speed is faster than the
AIA-inferred speed, which is indeed the case for the majority
of events (Fig.~18d, 19c, 19d). The slowest measured speeds 
amount to $v_{\mathrm{LE}} \gapprox 100$ km s$^{-1}$ for LASCO and
to $v_{\mathrm{BP}} \gapprox 30$ km s$^{-1}$ for AIA data. 
Earlier measurements with SOLWIND onboard P78-1 yield CME
speeds of  $v_{\mathrm{cme}} \approx 100-1500$ km s$^{-1}$
(Howard et al.~1985; Jackson and Howard 1993), 
or with LASCO $v_{\mathrm{cme}} \approx 100-1300$ km s$^{-1}$
(Chen et al.~2006).

There is also a time-dependence of the CME speed, but our
models (Fig.~4) indicate that the velocity function $v(t)$
reaches quickly an asymptotic value at coronal heights,
so that the speed is almost constant by the time it reaches
a distance of $x \gapprox 2.2$ solar radii, where we compare
LASCO/C2 and AIA-inferred CME speeds.

\subsection{	The Measurement of CME Masses 			} 

The CME masses determined with LASCO and AIA exhibit a large scatter
up to two orders of magnitude (Fig.~18b). We ask the question
what systematic errors affect the measurement of CME masses most?

For the white-light method, the mass is proportional to the
scattered polarized light (Appendix B, Eq.~B4). Summing up the
polarized brightness from a coronagraph, all the CME mass below
the coronagraph occulter height ($h \lapprox 2.2 R_{\odot}$) is
missing in the mass calculation (Fig.~1), which is most dramatically
underestimated during halo-CMEs, which propagate in the
direction to the observer. The missing mass was
estimated to be a factor of two for CMEs that are $\lapprox
40^\circ$ from the plane-of-sky (Vourlidas et al.~2010).
The LASCO CME catalog also lists many of the analyzed events
as ``poor'' or ``very poor events'', which probably characterizes 
the data noise and related detection sensitivity in the
polarized brightness difference images. The scatter plot
of CME masses shown in Fig.~18b indicates a low mass cutoff 
of $m \gapprox 0.3 \times 10^{15}$ g for AIA-inferred masses,
while the LASCO-inferred masses extend down to 
$m \gapprox 10^{13}$ g, which are likely to be underestimates
of the true masses. However, these underestimates do not
exclusively apply to halo-CMEs, because we were not able to
find a center-to-limb effect. Early studies, containing statistics 
of nearly 1000 CME events observed with the SOLWIND instrument
on the P78-1 satellite, exhibit a range of $m_{\mathrm{cme}} 
\approx 10^{14}-10^{17}$ g for CME masses
(Howard et al.~1985; Jackson and Howard 1993). 
A recent statistical study of 7741 CMEs detected with LASCO
shows a distribution with a range of 
$m \approx 10^{12}-10^{16}$ g, with the bulk of events in the range of
$m \approx 10^{14}-10^{15.5}$ g (Aarnio et al.~2011).

What are the advantages and caveats of EUV dimming-derived CME 
masses?  The EUV emission measure scales with the square of the
electron density, $EM \propto n_e^2 V$ (Eq.~11),
a property that already leads to a nonlinear amplification 
in the contrast of any detected EUV feature, in particular
to EUV brightenings and dimmings with respect to the
background EUV emission. Moreover, because no coronagraph
is needed, the entire CME source region is visible to an
EUV imager without any occultation effects (except behind
the solar limb, as it may happen for stealth CMEs).
The CME mass is then directly obtained from the volume
integral $m_{\mathrm{cme}}=n_e m_p L^2 \lambda$ (Eq.~13).
Some uncertainties in the CME mass determination may result
from the inhomogeneity of the mass distribution inside the
coronal region that corresponds to the CME initial source 
volume, which depends on the measurement of the projected
dimming area $A_p$, which we estimate to be a factor of
$\lapprox 2$ in the CME mass estimate, being a relatively
small correction for the observed variation of CME masses
by two orders of magnitude ($m \approx (0.3-30) \times 
10^{15}$ g). In comparison, earlier measurements with 
SOLWIND on the P78-1 satellite, exhibit CME masses of
$m_{\mathrm{cme}} \approx 10^{14}-10^{17}$ g.

\subsection{	The Measurement of CME Kinetic Energies 	} 

The kinetic energy, $E_{\mathrm{kin}}=(1/2) m_{\mathrm{cme}} v^2$ (Eq.~17), 
is a combination of the CME mass and CME speed, and thus is 
subject to all of their uncertainties discussed above, mostly
affected by the velocity, which has a square dependence.
The obtained kinetic energies show indeed a large scatter
between the LASCO and AIA-derived values, in the order of
$\approx 4$ orders of magnitude (Fig.~18f). The largest discrepancy
comes from the different definitions of leading edge $v_{\mathrm{LE}}$
and bulk plasma speeds $v_{\mathrm{BP}}$, which varies by about
1.5 orders of magnitude (Fig.~18d), and causes with its
square dependence a variation of 3 orders of magnitude in
the kinetic energy. The range of kinetic energies
obtained with both AIA and LASCO is $E_{\mathrm{kin}} \approx 
10^{28}-10^{33}$ erg (Fig.~18f).
Earlier studies, containing statistics of nearly 1000 CME 
events observed with the SOLWIND instrument on the P78-1 
satellite, exhibit a range of $E_{\mathrm{kin}} \approx 
10^{29}-10^{32}$ erg for CME kinetic energies (Howard 
et al.~1985; Jackson and Howard 1993). 

\subsection{	The Measurement of Gravitational Energies 	} 

The gravitational potential energy to lift a CME from the
bottom of the corona to infinity depends only on the CME mass
(Eq.~22). For solar gravity, the required escape velocity is
$v_{\mathrm{esc}} = 618$ km s$^{-1}$. Note that all CME
velocities we quote are close to the asymptotic limit at
large distances from the Sun, and thus represent roughly
the sum of the escape velocity and the observed moving
velocity. CMEs that are accelerated to a lower velocity
than the escape speed, $v_{\mathrm{init}} < v_{\mathrm{esc}}$, produce
so-called ``failed CMEs'', which initially move upward
and fall back after some time. In this study we do not
distinguish between ``failed/confined'' and ``eruptive''
CME events, assuming that radial adiabatic expansion occurs 
for both phenomena (at least during the initial ``explosive''
phase), and thus our data set of 399 M and X-class flare
events may contain both types, while the coincident 247
LASCO events are all eruptive events, since they are detected
at large distances $h \gapprox 2.2 R_{\odot}$. 

For an estimate the flare energy budget it is important
to compute both the kinetic and the gravitational potential 
energy for each CME event. For the 399 analyzed CME events
we find a range of $E_{\mathrm{kin}} = 3 \times 10^{27} - 10^{33}$ erg
for kinetic CME energies, and a range of 
$E_{\mathrm{grav}} = 2 \times 10^{29} - 6 \times 10^{31}$ erg (Table 1).
Actually, we find that the gravitational energy makes up
a fraction $E_{\mathrm{grav}}/(E_{\mathrm{tot}} = 0.75\pm0.28$
of the total CME energy $E_{\mathrm{tot}}=(E_{\mathrm{grav}}+E_{\mathrm{kin}})$
in the average. Only for the most
energetic CMEs the kinetic energy is larger than the gravitational
energy, which is the case for 22\% of the M and X-class flares.

In an earlier study on the global energetics of flares and CMEs,
the CME kinetic energy $E_{\mathrm{kin}}$ in the rest frame of the Sun, 
as well as in the solar wind frame were calculated, and the 
gravitational potential energy $E_{\mathrm{grav}}$ was calculated
(Emslie et al.~2004, 2005, 2012). The gravitational energy 
was found to be about an order of magnitude lower than the 
kinetic CME energy, in contrast to the study of Vourlidas 
et al.~(2000), where it was found that the center of mass
accelerates for most of the events, while the CMEs achieve
escape velocity at heights of around 8-10 $R_{\odot}$,
and that the potential energy is greater than the kinetic
energy for relatively slow CME events (which constitute
the majority of events).

\subsection{	Global Energetics of Flares and CMEs		} 

We are now in the position to study the global energetics of solar
flares and CMEs, by combining the dissipated magnetic energies
$E_{\mathrm{magn}}$ (Paper I), the available free (magnetic) energies
$E_{\mathrm{free}}$ (Paper I), the multi-thermal energies $E_{\mathrm{th}}$
(Paper II), the nonthermal energies $E_{\mathrm{nth}}$ (Paper III),
and the CME (kinetic and gravitational) energies from this study.
In Fig.~20 we present scatter plots between the different types
of energies: The LASCO-inferred CME kinetic energies (Fig.~20
left-hand panels) show a similar large scatter with the other
forms of energy as the AIA-inferred energies do (Fig.~20 middle 
panels). However, since the gravitational energy dominates
the kinetic energy for most CME events, we have to compare
the total (kinetic plus gravitational) CME energies, which
we show for the AIA data (Fig.~20 right-hand panels). 
Interestingly, the total CME energy exhibits a substantially
smaller scatter with other forms of energies, about a factor
$\approx 3$ less than the kinetic energy (see the factors
of the logarithmic standard deviations indicated in the top
right corner of each panel in Fig.~20), which substantially
improves the correlations between different forms of energy
that are dissipated and converted during flare/CME events.

For a test of the magnetic energy budget that can support 
a CME, the mean energy ratio is important. We find that the
ratio of the total CME energy to the dissipated magnetic
energy is (in the logarithmic mean) $q_E=0.07$, with a
standard deviation by a factor of 5.4 (Fig.~20c), which 
corresponds to a range of $q_e=0.01-0.38$. This is an 
important result that warrants that the dissipated
magnetic energy (as estimated from the untwisting of 
helically distorted field lines by vertical currents;
Paper I) is virtually always sufficient to support the 
launch of the observed CMEs. Also the available free
energy shows a similar ratio, $q_e=0.11$, with a
standard deviation by a factor of 6.3 (Fig.~20f), which 
corresponds to a range of $q_e=0.02-0.69$. In principle,
the free energy should be larger than the actually
dissipated magnetic energy, and thus the difference
must result from a combination of measurement uncertainties.  

A most striking result is the relationship between total
CME energies and thermal flare energies, which shows a
much reduced scatter down to a factor of 3.5, which
reveals a true correlation between these two forms of
energies. The (logarithmic) mean ratio is
$q_e=0.77$, within a factor of 3.5 (Fig.~20i).
In other words there is about the same amount of energy
that goes into a CME as is converted into thermal flare
energy (as measured in soft X-rays). We find a similar 
result for the relationship between total CME energies
and nonthermal energies (Fig.~20l), calculated from
the thick-target model with a low-energy cutoff of 20 keV.

If we envision a magnetic reconnection scenario with vertical 
eruption, such as described by Carmichael (1964), Sturrock
(1996), Hirayama (1974), Kopp and Pneuman (1976),
Tsuneta 1996, 1997, Shibata (1995), the plasma heating
by precipitating electrons and subsequent chromospheric
evaporation is a direct consequence of the accelerated
electrons in the reconnection region, behind a rising
prominence and CME, which explains the correlation 
between (multi-)thermal energies and total CME energies.
In a previous study of 37 impulsive flare/CME events it was found that the
CME peak velocity is highly correlated with the total energy in
nonthermal electrons, which supports the idea that the acceleration
of the CME and particle acceleration in the associated flare
have a common energy source, most likely magnetic reconnection
in the wake of an eruptive CME (Berkebile-Stoiser et al.~2012). 

\section{		CONCLUSIONS			}

The phenomenon of coronal mass ejections can be detected in two
major wavelength regimes, either in white-light as it has been
traditionally done with coronagraphs (e.g., with SOLWIND, LASCO/SOHO,
COR-2/STEREO) or in extreme ultra-violet images (e.g., with
EIT/SOHO, EUVI/STEREO, AIA/SDO). The two methods are truly complementary
and are based on different physical emission mechanisms. The white-light
method relies on Thomson scattering of white light on CME particles
that produces a polarized brightness in coronagraph images, and can
only be imaged a few solar radii away from the Sun due to the
necessity of an occulter disk that eclipses the (unpolarized) bright 
photospheric light. The EUV dimming method, in contrast, detects
first an increase of the EUV emission measure in the CME source region
due to chromospheric evaporation in flare loops, followed by a rapid
decrease in the EUV brightness due to the adiabatic expansion of the
compressed flare plasma out into the heliosphere along an open
magnetic field channel. While the white-light method has been
used quantitatively since about four decades, the EUV dimming method 
has been discussed and interpreted since two decades, mostly in a 
qualitative manner, but in this study we develop a simple EUV dimming
model that is fitted to a large statistical data set of solar 
flare/CME events in a quantitative way, for the first time. 
This work is part of a larger project on the global energetics
of solar flare and CME events, encompassing all GOES M and X-class
flares detected with AIA/SDO during the first 3.5 years of its mission,
amounting to 399 events. The major conclusions of this study are:

\begin{enumerate}
\item{\underbar{The EUV Dimming Model:} consists of a geometric
volume model $V(t)$ that takes the gravitational stratification into 
account (with a density scale height $\lambda$ corresponding to the 
pre-flare temperature) and corrects for center-to-limb projection 
effects, evaluates the mean electron density and total dimming mass
from differential emission measure distributions in the temperature
range of $T_e \approx 0.5-20$ MK (provided by AIA with the 
spatial-synthesis DEM method), fits the total EUV emission measure 
profile $EM(t)$ in the dimming region with a dynamic model based on 
radial adiabatic expansion, $EM(t) \propto V(t)^{-1}$, and infers the
height-time profile $x(t)$, velocity $v(t)$, and acceleration
$a(t)$ of the CME bulk plasma by fitting the EUV dimming profile
$EM(t)$ to the observed emission measure evolution. This model
takes full advantage of the AIA density and temperature diagnostics
and can quantitatively describe the evolution of all (macroscopic)
CME parameters in the CME source region and during the propagation
through the corona over a range of a few density scale heights,
which complements the white-light method that probes the evolution
of the CME in the heliosphere at distances of 
$x(t) \gapprox 2 R_{\odot}$ from Sun center. The performance of
the automated forward-fitting code can be considered as satisfactory,
given the fact that a unique best-fit solution is found in all 399
flare events and a goodness-of-fit distribution with a mean and 
standard deviation of $\chi = 1.6\pm1.1$. The detection efficiency
of the EUV dimming method is 100\% for M and X-class flares and a
relatively high dimming ratio of $q_{\mathrm{dimm}}=0.82\pm0.12$ is found.}

\item{\underbar{CME speeds - Simple and Complex Dimming Events:} 
Simple events
show a rapid initial increase of the emission measure that coincides
with the flare (GOES) peak time within $\Delta t = 5 \pm 7$ min,
followed by a rapid dimming down to almost the pre-flare level,
closely following the predicted time profile for radial adiabatic 
expansion, $EM(t) \propto V(t)^{-1}$. These simple events yield
expansion speeds of $v \approx 100-3000$ km s$^{-1}$ similar to
the CME leading edge speeds obtained with the white-light method.
On the other extreme we find complex events, mostly in flares with
long durations (with GOES flare durations of $D \approx 0.5-4.0$ hrs),
which appear to consist of multiple expansion phases in space and
time. The best fits of our single-expansion model may be hampered
by a temporal and spatial convolution bias in these complex events,
which tend to yield lower (time-averaged) expansion speeds down
to the range of $v \approx 30-100$ km s$^{-1}$, which are below
the typical speeds measured with the white-light method. We interpret
these slow speeds in terms of CME bulk plasma speeds, which are 
expected to be lower than the CME leading edge speeds by definition. A
consequence of the spatio-temporal convolution bias is a reciprocal
correlation of the bulk plasma speed with the flare duration,
$v \propto D^{-1}$.}

\item{\underbar{CME masses:} The EUV dimming method (using AIA data)
yields CME masses in the range of $m_{\mathrm{cme}} = (0.1 - 30) 
\times 10^{15}$ g, which is consistent with CME masses inferred with 
the white-light method (using SOLWIND, EIT, or LASCO data). The
EUV dimming method is very sensitive because the EUV emission measure
depends on the square of the electron density, $EM \propto n_e^2 V$,
which yields a high contrast for every density enhancement, while
the white-light method is only linearly sensitive to the density.
This may explain, why the fraction of LASCO-detected CMEs is only
30\%-70\% (depending on the simultaneity requirement), while AIA
has a 100\% detection efficiency for M and X-class flares and
shows near-simultaneity with the GOES flare peak time within
a few minutes. Low masses
detected with LASCO in the range of $10^{13}-10^{14}$ g may be
underestimates of the true CME masses, since such low values are
not found with AIA. The EUV dimming method is equally sensitive
to limb and halo CMEs, while the white-light method systematically
underestimates the mass of halo CMEs due to the coronagraph 
occultation.}

\item{\underbar{CME energies:} The EUV dimming method yields
CME kinetic energies in the range of 
$E_{\mathrm{kin}} = (0.003 ... 960)\times 10^{30}$ erg,
CME gravitational potential energies of 
$E_{\mathrm{grav}} = (0.2 ... 57)\times 10^{30}$ erg,
and CME total energies of $E_{\mathrm{tot}} = E_{\mathrm{grav}} + E_{\mathrm{kin}} 
= (0.25 ... 1000)\times 10^{30}$ erg.
The gravitational potential energy has an average fraction of
$E_{\mathrm{grav}}/E_{\mathrm{tot}} = 0.75\pm0.28$, while
only 22\% (consisting of the most energetic M and X-class flares)
have a kinetic energy in excess of the gravitational potential energy.
It is therefore imperative to include the gravitational energy
in the overall energy requirement of CMEs. 
This is consistent with the finding of
an earlier analysis by Vourlidas et al.~(2000), but is apparently 
different from the energetics study of Emslie et al.~(2012),
probably because of a bias for the most energetic (mostly
X-class) eruptive events in that sample.}

\item{\underbar{Global Energetics of Flares and CMEs:} Comparing
the CME kinetic energy with other forms of flare energies reveals
a huge scatter, for both LASCO and AIA-inferred CME energies, up
to $\pm 2$ orders of magnitude, which is mostly caused by the quadratic
dependence on the CME speed, which can differ substantially for
CME leading edge or CME bulk plasma speeds. However, comparing
the CME total energy, by combining the gravitational and kinetic
energy, we find a much smaller scatter, down to a factor of 3.5
for the ratio of the CME total energy to the multi-thermal energy.
Regarding the overall energy partition in flares we find that the
CME total energy is virtually always lower than the dissipated
(or the free) magnetic energy, by a fraction of 
$E_{\mathrm{cme}}/E_{\mathrm{mag}} \approx 1\%-40\%$
of the dissipated nonpotential magnetic energy. 
We find that the amount of energy that goes into (multi-)thermal 
plasma heating (with a logarithmic mean of $E_{\mathrm{th}}/E_{\mathrm{cme}}=0.77$), 
or the nonthermal energy of accelerated electrons 
($E_{\mathrm{nth}}/E_{\mathrm{cme}}=0.72$), are about equal to the amount that goes
into the production of a CME, which is consistent with a magnetic 
origin of CMEs and with the thick-target bremsstrahlung model for
particle acceleration.} 

\item{\underbar{Self-Organized Criticality (SOC):} For most CME parameters
($L, A, A_p, V, EM, v, E_{\mathrm{kin}}, E_{\mathrm{grav}}, E_{\mathrm{tot}}$) we find occurrence
frequency distributions with a power law tail, which has been
interpreted as evidence for self-organized criticality systems,
first by Lu and Hamilton (1991) for the case of solar flares
(for a recent review of SOC in solar physics and astrophysics
see Aschwanden et al.~2016).
The power law distributions have all different slopes for each
CME parameter, such as $N(E_{\mathrm{kin}}) \propto E_{\mathrm{kin}}^{-2.2\pm0.2}$ for
the CME kinetic energy, or $N(E_{\mathrm{tot}}) \propto E_{\mathrm{tot}}^{-1.9\pm0.3}$
for the CME total energy. A new result is that we can correctly
predict the specific power law slopes for all CME parameters with
a standard SOC model based on fractal-diffusive avalanching
of a slowly-driven SOC system (Aschwanden 2012), if we 
adopt two modifications, namely the CME source volume scaling
$V \propto L^2$ in a gravitationally stratified atmosphere, and
the convolution bias of adiabatic expansion velocities, 
$v \propto D^{-1}$, which scales reciprocally with the flare 
duration $D$.}

\end{enumerate}

What problems are left for the future? Although we constructed an EUV
dimming model that fits all analyzed flare events in our AIA data set, 
there are some caveats and issues that merit further investigations: 
Can the ``complex'' events be modeled with a superposition of
spatially and temporally separated ``simple'' event evolutions?
How can the CME leading edge speeds and CME bulk plasma speeds 
be unified into a single dynamic model that would make EUV and
white light-inferred speeds compatible? How can the timing of the
dimming with respect to the flare peak times be reconciled in EUV
and white-light data, yielding a unique discrimination of which CME
detected by LASCO corresponds to a particular flare detected by
GOES or AIA? How can the correlation between CME energies and
other forms of (magnetic, thermal, nonthermal) energies be improved
with more accurate models? Does the EUV dimming model based on
radial adiabatic expansion fit weaker (GOES C-class) flares and how
does it differ for failed/confined CEMs and eruptive CMEs?
How can the spatio-temporal information of the CME source region
be characterized (with fractals, multi-fractals, fractal diffusion)
and is there a scaling law between spatial and temporal parameters?
How can the pre-dimming increase in emission measure be modeled?
What information on the physical process of a CME creation and
the Lorentz forces of the magnetic field acting in the CME source
region can be extracted from the combined AIA and LASCO data.
Ultimately we aim for a deeper understanding of the physical processes
that govern plasma instabilities in the creation of a CME.

\bigskip
\acknowledgements
We acknowledge useful comments and discussions with Gordon Emslie,
Nat Gopalswamy, Nariaki Nitta, Manuela Temmer, Barbara Thompson,
Astrid Veronig, Angelos Vourlidas, and Jie Zhang.
This work was partially supported by NASA contract NNG 04EA00C 
of the SDO/AIA instrument.

\section*{APPENDIX A: CME Kinematics}

To model the kinematics of a CME during its propagation in the solar corona 
we explore here four simple acceleration functions: a constant (Model 1), 
a linearly decreasing (Model 2) , a quadratically decreasing (Model 3),
and an exponentially decreasing function (Model 4):
$$
        a(t) = \left\{ \begin{array}{lll}
        a_0                      		& \mbox{for $t_0 < t < t_A$}        & {\rm (Model\ 1)} \\
        a_0 (1 - (t-t_0)/\tau)   		& \mbox{for $t_0 < t < t_A$}        & {\rm (Model\ 2)} \\
        a_0 (1 - (t-t_0)/\tau)^2 		& \mbox{for $t_0 < t < t_A$}        & {\rm (Model\ 3)} \\
        a_0 \exp{[-(t-t_0)/\tau]}		& \mbox{for $t_0 < t < t_A$}        & {\rm (Model\ 4)}
        \end{array}
        \right.
\eqno(A1)
$$
The time interval of acceleration is defined by $\tau=t_A-t_0$ in all models,
while no acceleration occurs before $t < t_0$ or after $t > t_A$. 
A graphical representation of these 4 acceleration profiles is shown in Fig.~4a.

From the acceleration profile $a(t)$ (parameterized with 3 variables $t_0$, $t_A$,
$a_0$) we can then directly calculate the velocity profile $v(t)$ by time
integration, which yields during the acceleration phase ($t_0 \le t \le t_A$),
$$
        v(t) = \int_{t_0}^t a(t) \ dt =
        \left\{ \begin{array}{lll}
        a_0 (t-t_0)        			& \mbox{for $t_0 < t < t_A$}      & {\rm (Model\ 1)} \\
        a_0 [(t-t_0 - (t-t_0)^2/2\tau ]    	& \mbox{for $t_0 < t < t_A$}      & {\rm (Model\ 2)} \\
        (1/3) a_0 \tau [1 - (1-(t-t_0)/\tau)^3]  & \mbox{for $t_0 < t < t_A$}      & {\rm (Model\ 3)} \\
        a_0 \tau [ 1 - \exp(-(t-t_0)/\tau) ]    & \mbox{for $t_0 < t < t_A$}      & {\rm (Model\ 4)} \\
        \end{array}
        \right.
\eqno(A2)
$$
and at times after the acceleration phase $(t > t_A)$,
$$
        v(t) = \int_{t_0}^t a(t) \ dt =
        \left\{ \begin{array}{lll}
        a_0 \tau           			& \mbox{for $t > t_A$}      & {\rm (Model\ 1)} \\
        (1/2) a_0 \tau 				& \mbox{for $t > t_A$}      & {\rm (Model\ 2)} \\
        (1/3) a_0 \tau      	   		& \mbox{for $t > t_A$}      & {\rm (Model\ 3)} \\
        a_0 \tau [ 1 - \exp(-(t-t_0)/\tau) ]    & \mbox{for $t > t_A$}      & {\rm (Model\ 4)} \\
        \end{array}
        \right.
\eqno(A3)
$$
Similarly we can derive the height-time profile $x(t)$ 
of the CME bulk mass by time integration of the velocity $v(t)$,
$$
        x(t) = \int_{t_0}^t v(t) \ dt =
        \left\{ \begin{array}{lll}
        x_0 +(1/2) a_0 (t-t_0)^2                & \mbox{for $t_0 < t < t_A$} & {\rm (Model\ 1)} \\
        x_0 +(1/3) a_0 \tau^2              	& \mbox{for $t_0 < t < t_A$} & {\rm (Model\ 2)} \\
        x_0+(1/3)a_0 \tau [(t-t_0)+(1/4) \tau [(1-(t-t_0)/\tau]^4-1 ] & \mbox{for $t_0 < t < t_A$} & {\rm (Model\ 3}) \\
        x_0 + a_0 \tau (t-t_0) - a_0 \tau^2 [ 1 - \exp(-(t-t_0)/\tau ] & \mbox{for $t_0 < t < t_A$} & {\rm (Model\ 4}) \\
        \end{array}
        \right.
\eqno(A4)
$$
and at times after the acceleration phase $(t > t_A)$,
$$
        x(t) = \int_{t_0}^t v(t) \ dt =
        \left\{ \begin{array}{lll}
        x_0+(1/2) a_0 \tau^2 + a_0 \tau (t-t_A) & \mbox{for $t > t_A$} & {\rm (Model\ 1)} \\
        x_0+(1/3)a_0 \tau^2 + (1/2) a_0 \tau (t-t_A) & \mbox{for $t > t_A$} & {\rm (Model\ 2)} \\
        x_0 + (1/4) a_0 \tau^2 + (1/3) a_0 \tau (t-t_A) & \mbox{for $t > t_A$} & {\rm (Model\ 3)} \\
        x_0 + a_0 \tau (t-t_0) - a_0 \tau^2 [ 1 - \exp(-(t-t_0)/\tau ]	& \mbox{for $t > t_A$} & {\rm (Model\ 4)} \\
        \end{array}
        \right.
\eqno(A5)
$$
where we allow for an integration constant $x_0$ that represents the initial height
of the CME bulk mass. 

\section*{ APPENDIX B : THE WHITE-LIGHT CORONAGRAPH METHOD      }

The method of extracting CME masses from white-light
observations of Thomson scattering, detected in form of
polarized brightness in coronagraph data, originated from
Minnaert (1930), van de Hulst (1950), Billings 1966, and
was further developed by Vourlidas and Howard (2006).
White light undergoes Thomson scattering in the solar corona,
which is sensitive to the geometry of the distribution of
scattering particles and the direction to the observer.
The scattering cross-section
depends on the angle $\chi$ between the line-of-sight
and the radial direction through the scattering electron as,
$$
        {d\sigma \over d\omega}={1 \over 2} r_e^2 (1 + \cos^2{\chi}) \, ,
	\eqno(B1)
$$
where ${d\sigma / d\omega}$ is the differential cross-section in units of
[$cm^2\ sr^{-1}$] and
$r_e=e^2/m_e c^2=2.82 \times 10^{-13}$ cm is the classical electron
radius. By integrating over all solid angles we obtain the total
cross-section for perpendicular scattering, the so-called \emph{Thompson
cross-section} for electrons,
$$
        \sigma_T={8 \pi \over 3} r_e^2 = 6.65 \times 10^{-25} \  {\rm cm}^2 \,.
\eqno(B2)
$$
The total scattered radiation $I(x,y)$ can then be calculated by integrating
over the source locations of the photons (the photosphere) and the
scattering electrons (with a 3D distribution $n_e(x,y,z)$) along the
line-of-sight $z$, as a function of the scattering angle $\chi(x,y,z)$
with respect to the observers line-of-sight.  The degree of polarization $p$,
which is observed in the \emph{polarized brightness (pB)} component of
white-light images, is defined as,
$$
        p = {I_T - I_R \over I_T + I_R} = {I_P \over I_{\mathrm{tot}}} \, ,
\eqno(B3)
$$
where $I_T$ and $I_R$ represent the tangential and radial terms of the
total scattered radiation. Many coronagraphs (such as those on STEREO) have
the capability to measure linear polarization in three orientations,
from which the total $I_{\mathrm{tot}}$ and polarized brightness $I_P$ can be
derived. The total scattered radiation is proportional
to the electron density $n_e(x,y,z)$ of the scattering corona,
in contrast to the square-dependence of the observed brightness
(or emission measure)
on the density (Eq.~11) for free-free emission in soft X-rays and EUV.
Since the differential cross section $d\sigma / d\omega$ (Eq.~B1)
varies only by a factor of two with angle $\chi$, the plane-of-the-sky
or plane-of-maximum-scattering approximations is poor. It is essential
for any tomography approach that aims to reconstruct an extended density
distribution to treat the observations as the result of an extended integral
along the line-of-sight. The plane of maximum scattering has often been
approximated with the plane-of-sky in the past,
which is appropriate for locations near the solar limb, but needs to
be corrected with the actual plane-of-maximum-scattering, for
sources at large distances from the Sun (Vourlidas and Howard 2006).

Masses of CMEs have traditionally been determined from coronagraph
observations in white light. The CME mass is estimated by subtracting
a pre-event image and assuming that the remaining excess brightness
$I_{\mathrm{obs}}$ is due to Thomson scattering $I_e(\vartheta)$ by electrons
at an angle $\vartheta$
from the plane of the sky (Minnaert 1930; van de Hulst 1950; Jackson 1962;
Billings 1966; Poland et al.~1981; Vourlidas \& Howard 2006).
Assuming a standard abundance
of fully-ionized hydrogen with 10\% helium, the CME mass is
$$
        m_{\mathrm{cme}} = {I_{obs} \over I_e(\vartheta)} 2 \times 10^{-24}\ (g) \ .
\eqno(B4)
$$
Usually the calculations assume $\vartheta=0$ (all electrons are in
the plane of sky), which leads to a minimum value of the CME mass.
The CME mass is typically underestimated by a factor of 2 for most cases.
The masses of CMEs were measured with Skylab for 21 events in the range
of $m_{\mathrm{cme}}\approx (0.8-7.9) \times 10^{15}$ g (Jackson \& Hildner 1978),
with the Solwind coronagraph during 1979-1981 with an average of
$m_{\mathrm{cme}}=4.1 \times 10^{15}$ g (Howard et al. 1985), and with the
SOHO/LASCO coronagraph with an average of $m_{\mathrm{cme}}=1.7 \times 10^{15}$ g
(Vourlidas et al.~2000, 2002), covering a range of
$m_{\mathrm{cme}} \approx 10^{15}-10^{16}$ g (Subramanian \& Vourlidas 2007).
Only recently, when STEREO coronagraph observations became
available, the 3D propagation direction could be pinpointed and an
improved CME mass could be evaluated (Colaninno \& Vourlidas 2009).
Improved CME masses in the order of $m_{\mathrm{cme}} \approx (2.6-7.7) \times 10^{15}$ g
were determined from the STEREO/A and B COR2 instruments. The CME is detected
in the COR2 field-of-view about 1-4 hours later after launch from the
coronal base (Fig.~1), and the improved mass was found to increase with time,
asymptotically
approaching a constant value after the CME mass emerged out of the
occulted area at $r \le 4 R_{\odot}$, and thus verifying the original
LASCO observations (Vourlidas et al.~2000). 


\clearpage

\begin{deluxetable}{llll}
\tablecaption{Ranges and distributions of CME parameters 
measured in 399 CME events with AIA. The mean and standard deviations refer to the
slopes $p$ of power law distributions, or to the Gaussian normal distributions 
$x\pm \sigma_x$. The dimming delay is defined as the time difference 
$\Delta t = (t_{\mathrm{dimm}}-t_{\mathrm{peak}})$ between the beginning of EUV dimming 
and the GOES (1-8 \ang ) flare peak time.}
\tablewidth{0pt}
\tablehead{
\colhead{Parameter}&
\colhead{Range}&
\colhead{Distribution}&
\colhead{Mean and}\\
\colhead{}&
\colhead{}&
\colhead{type}&
\colhead{standard}\\
\colhead{}&
\colhead{}&
\colhead{}&
\colhead{deviation}}
\startdata
Length scale $L$	  & $(28 ... 361)$ Mm                     & power law & $p=-3.4\pm1.0$ \\ 
Projected area $A_p$      & $(0.2 ... 4.0)\times 10^{20}$ cm$^2$  & power law & $p=-1.3\pm0.1$ \\
CME dimming area $A$      & $(0.1 ... 13.0)\times 10^{20}$ cm$^2$ & power law & $p=-1.3\pm0.1$ \\
CME dimming volume $V$ 	  & $(0.06 ... 12)\times 10^{30}$ cm$^3$  & power law & $p=-2.5\pm0.6$ \\
Dimming time $\tau_{\mathrm{dimm}}$& $(0.5 ... 33)$ min           & power law & $p=-2.5\pm0.6$ \\
Propagation time $\tau_{\mathrm{prop}}$& $(0.3 ... 54)$ min       & power law & $p=-1.5\pm0.2$ \\
GOES flare duration	  & $(0.1 ... 7.0)$ hrs                   & power law & $p=-2.0\pm0.2$ \\
CME mass $m$		  & $(0.1 ... 30)\times 10^{15}$ erg      & power law & $p=-2.2\pm0.5$ \\
CME emission measure $EM$ & $(2.7 ... 334)\times 10^{27}$ cm$^{-3}$&power law & $p=-2.4\pm0.5$ \\
CME speed $v$             & $(26 ... 4100)$ km s$^{-1}$           & power law & $p=-1.9\pm0.3$ \\
CME kinetic energy $E_{\mathrm{kin}}$ & $(0.003 ... 960)\times 10^{30}$ erg & power law & $p=-1.4\pm0.1$ \\
CME grav. energy $E_{\mathrm{grav}}$ & $(0.2 ... 57)\times 10^{30}$ erg & power law & $p=-2.0\pm0.4$ \\
CME total energy $E_{\mathrm{tot}}$& $(0.25 ... 1000)\times 10^{30}$ erg   & power law & $p=-2.0\pm0.3$ \\
Dimming delay $\Delta \tau_{\mathrm{dimm}}$ & $\Delta \tau_{\mathrm{dimm}}=(-7 ... 108)$ min & Gaussian & $\Delta \tau=5.8\pm7.3$ min \\
pre-CME density $n_e$     & $(0.6 ... 6.8)\times 10^9$ cm$^{-3}$  & Gaussian  & $n_e=(1.3\pm0.6)\times 10^9$ cm$^{-3}$\\
pre-CME temperature $T_e$ & $(1.2 ... 5.0)$ MK                    & Gaussian  & $T_e=1.8\pm0.3$ MK \\
Density scale height $\lambda$ & $\lambda=(47 ... 235)$ Mm        & Gaussian  & $\lambda=86\pm12$ Mm \\
Dimming fraction $q_{\mathrm{dimm}}$ & $q_{\mathrm{dimm}}=(0.3 ... 1.0)$  & Gaussian  & $q_{\mathrm{dimm}}=0.82\pm0.12$ \\
Goodness-of-fit $\chi$    & $(0.06 ... 6.1)$                      & Gaussian  & $\chi=1.6\pm1.1$ \\
\enddata
\end{deluxetable}
\clearpage

\begin{deluxetable}{lllll}
\tablecaption{Predicted and observed occurrence frequency distributions $N(x)$ 
and correlations $x \propto y^p$ of CME parameters.}
\tablewidth{0pt}
\tablehead{
\colhead{Parameter}&
\colhead{Function}&
\colhead{Theoretical}&
\colhead{Observed}&
\colhead{Reference}\\
\colhead{}&
\colhead{}&
\colhead{Exponent}&
\colhead{Exponent}&
\colhead{Figure}}
\startdata
Length scale $L$	& $N(L) \propto L^d $	&  $d=-3.0$ 	& $d=-3.4 \pm1.0 $ & (Fig.~16a) \\
CME volume $V$ 		& $ V   \propto L^a $	&  $a=-2.0$	& $a=-1.98\pm0.02$ & (Fig.~14c) \\
			& $N(V) \propto V^p $   &  $p=-2.2$     & $p=-2.2 \pm0.5 $ & (Fig.~16b) \\ 
CME mass $m$		& $ m   \propto V^a $   &  $a=+1.0$     & $a=+1.07\pm0.06$ & (Fig.~14g) \\
			& $N(m) \propto m^p $   &  $p=-2.2$     & $p=-2.0 \pm0.4 $ & (Fig.~16f) \\
Emission measure $EM$   & $ EM  \propto V^a $   &  $a=+1.0$     & $a=+0.9 \pm0.4 $ & (Fig.~14d) \\
		        & $N(EM)\propto EM^p$   &  $p=-2.2$     & $p=-2.4 \pm0.5 $ & (Fig.~16d) \\
Duration $D$            & $ D   \propto L^a $   &  $a=+2.0$     & $....$           &            \\
			& $N(D) \propto D^p $   &  $p=-2.2$     & $p=-2.4 \pm0.6 $ & (Fig.~16c) \\
CME speed $v$           & $ v   \propto D^a $   &  $a=-1.0$     & $p=-1.6 \pm0.6 $ & (Fig.~15g) \\
			& $N(v) \propto v^p $   &  $p=-1.8$     & $p=-1.9 \pm0.2 $ & (Fig.~16e) \\ 
CME kinetic energy $E_{\mathrm{kin}}$ & $E_{\mathrm{kin}}   \propto v^a $   &  $a=+2.0$     & $p=+2.2 \pm0.2 $ & (Fig.~14h) \\
			& $N(Ek)\propto Ek^p$   &  $p=-1.4$     & $p=-1.4 \pm0.1 $ & (Fig.~16g) \\
CME grav. energy $E_{\mathrm{grav}}g$   & $E_{\mathrm{grav}}\propto m^a $   &  $a=+1.0$     &                  &            \\
			& $N(Eg)\propto Eg^p$   &  $p=-2.2$     & $p=-2.0 \pm0.4 $ & (Fig.~16h) \\
CME total energy $E_{\mathrm{tot}}t$   & $E_{\mathrm{tot}} \propto Ek+Eg$  &               &                  &            \\
			& $N(Et)\propto Et^p$   &  $p=-1.9$     & $p=-2.0 \pm0.3 $ & (Fig.~16i) \\
\enddata
\end{deluxetable}
\clearpage

\begin{deluxetable}{rrrrrrrrrrrr}
\footnotesize
\tablecaption{Temporal and spatial parameters of the first 10 entries 
(out of the 399 events) listed in the complete machine-readable data file.
The start time $t_{\mathrm{start}}$ refers to the GOES flare catalog, 
the rise time is $\tau_{\mathrm{rise}}=t_{\mathrm{peak}}-t_{\mathrm{start}}$, 
the decay time is $\tau_{\mathrm{decay}}=t_{\mathrm{end}}-t_{\mathrm{peak}}$, 
the dimming delay is $\Delta t_{\mathrm{dimm}}=t_{\mathrm{dimm}}-t_{\mathrm{peak}}$, 
the dimming duration is $\tau_{\mathrm{dimm}}=t_{\mathrm{half}}-t_{\mathrm{dimm}}$, 
the propagation time is $\tau_{\mathrm{prop}}=\lambda/v$.}
\tablewidth{0pt}
\tablehead{
\colhead{$\#$}&
\colhead{GOES}&
\colhead{Helio-}&
\colhead{Date}&
\colhead{Start}&
\colhead{Rise}&
\colhead{Decay}&
\colhead{Dimm.}&
\colhead{Dimm.}&
\colhead{Prop.}&
\colhead{Length}&
\colhead{Projected}\\
\colhead{}&
\colhead{class}&
\colhead{graphic}&
\colhead{}&
\colhead{time}&
\colhead{time}&
\colhead{time}&
\colhead{delay}&
\colhead{duration}&
\colhead{time}&
\colhead{scale}&
\colhead{area}\\
\colhead{}&
\colhead{}&
\colhead{position}&
\colhead{[UT]}&
\colhead{[s]}&
\colhead{[s]}&
\colhead{[s]}&
\colhead{[s]}&
\colhead{[s]}&
\colhead{[s]}&
\colhead{[cm]}&
\colhead{[cm$^2$]}}
\startdata
   1 &M2.0 &N23W47 &2010-06-12 &00:30 &1680 & 240 & 240 &  71 &  80 &1.12E+10 &1.44E+20 \\
   2 &M1.0 &S24W82 &2010-06-13 &05:30 & 540 & 300 & 300 & 141 & 223 &2.09E+10 &1.94E+20 \\
   3 &M1.0 &N13E34 &2010-08-07 &17:55 &1740 &1380 &1020 & 426 & 908 &1.73E+10 &3.19E+20 \\
   4 &M2.9 &S18W26 &2010-10-16 &19:07 & 300 & 180 &  60 & 135 &  50 &8.10E+09 &9.24E+19 \\
   5 &M1.6 &S20E85 &2010-11-04 &23:30 &1680 & 840 & 240 & 314 & 445 &8.60E+09 &7.81E+19 \\
   6 &M1.0 &S20E75 &2010-11-05 &12:43 &2760 &2220 & 240 & 498 & 839 &9.50E+09 &9.90E+19 \\
   7 &M5.4 &S20E58 &2010-11-06 &15:27 & 540 & 480 & 420 &  60 & 497 &1.11E+10 &1.43E+20 \\
   8 &M1.3 &N16W88 &2011-01-28 &00:44 &1140 & 420 & -60 & 245 & 239 &1.26E+10 &9.96E+19 \\
   9 &M1.9 &N16W70 &2011-02-09 &01:23 & 480 & 240 & 240 & 205 & 216 &1.08E+10 &1.24E+20 \\
  10 &M6.6 &S21E04 &2011-02-13 &17:28 & 600 & 540 & 240 & 234 & 362 &1.67E+10 &3.00E+20 \\
\enddata
\end{deluxetable}

\begin{deluxetable}{rrrrrrrrr}
\footnotesize
\tablecaption{CME parameters of the first 10 entries 
(out of the 399 events) listed in the complete machine-readable data file.
The total emission measure $EM_{\mathrm{tot}}$ is measured at the peak time of the
emission measure profile, the electron density $n_e$ and temperature $T_e$
refer to the flare start time, the CME velocity $v$, kinetic energy $E_{\mathrm{kin}}$
and gravitational potential energy $E_{\mathrm{grav}}$ refer to the asymptotic values
at the end time of the analyzed time interval.}
\tablewidth{0pt}
\tablehead{
\colhead{$\#$}&
\colhead{Dimming}&
\colhead{Emission}&
\colhead{Electron}&
\colhead{Electron}&
\colhead{CME}&
\colhead{CME}&
\colhead{Kinetic}&
\colhead{Gravitational}\\
\colhead{}&
\colhead{ratio}&
\colhead{measure}&
\colhead{density}&
\colhead{temperature}&
\colhead{mass}&
\colhead{velocity}&
\colhead{energy}&
\colhead{energy}\\
\colhead{}&
\colhead{}&
\colhead{$EM_{\mathrm{tot}}$}&
\colhead{$n_e$}&
\colhead{$T_e$}&
\colhead{$m$}&
\colhead{$v$}&
\colhead{$E_{\mathrm{kin}}$}&
\colhead{$E_{\mathrm{grav}}$}\\
\colhead{}&
\colhead{}&
\colhead{[cm$^{-3}$]}&
\colhead{[cm$^{-3}$]}&
\colhead{[K]}&
\colhead{[g]}&
\colhead{[cm s$^{-1}$]}&
\colhead{[erg]}&
\colhead{[erg]}}
\startdata
   1 &0.93 &7.89E+27 &1.10E+09 &1.60E+06 &1.72E+15 &9.29E+07 &7.44E+30 &3.28E+30 \\
   2 &0.79 &5.99E+27 &6.31E+08 &1.40E+06 &3.07E+15 &2.96E+07 &1.35E+30 &5.86E+30 \\
   3 &0.78 &1.15E+28 &1.29E+09 &1.60E+06 &4.79E+15 &8.20E+06 &1.61E+29 &9.14E+30 \\
   4 &0.95 &1.51E+28 &1.58E+09 &1.80E+06 &1.47E+15 &1.65E+08 &2.00E+31 &2.80E+30 \\
   5 &0.85 &1.04E+28 &1.15E+09 &1.80E+06 &1.19E+15 &1.87E+07 &2.09E+29 &2.27E+30 \\
   6 &0.86 &8.88E+27 &1.07E+09 &1.80E+06 &1.36E+15 &9.90E+06 &6.70E+28 &2.60E+30 \\
   7 &0.58 &2.38E+28 &1.82E+09 &1.80E+06 &3.14E+15 &1.68E+07 &4.43E+29 &5.98E+30 \\
   8 &0.95 &5.04E+27 &6.46E+08 &1.60E+06 &1.29E+15 &3.11E+07 &6.27E+29 &2.46E+30 \\
   9 &0.89 &6.50E+27 &9.12E+08 &1.80E+06 &1.49E+15 &3.86E+07 &1.11E+30 &2.84E+30 \\
  10 &0.94 &2.37E+28 &1.95E+09 &1.40E+06 &6.06E+15 &1.83E+07 &1.02E+30 &1.16E+31 \\
\enddata
\end{deluxetable}
\clearpage


\begin{figure}
\plotone{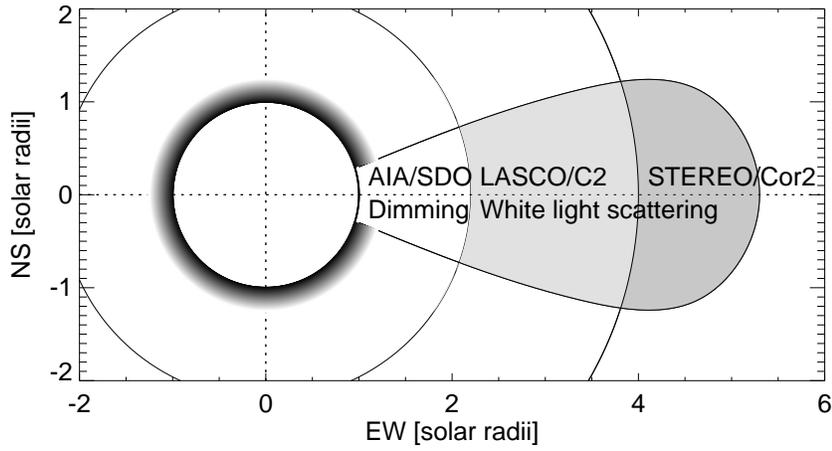}
\caption{Schematic of CME mass determination methods in EUV vs. white light.
Using the EUV method, the CME mass is calculated from the missing mass
that causes the EUV dimming in the lower corona at $r \lapprox 1.1 R_{\odot}$,
while the white-light method measures the excess of brightness in a coronagraph image,
e.g., at $r \gapprox 2.2$ (or 4.0) $R_{\odot}$ beyond the occulting disk of
the SOHO/LASCO/C2 (or STEREO/COR2) coronagraphs.}
\end{figure}

\begin{figure}
\plotone{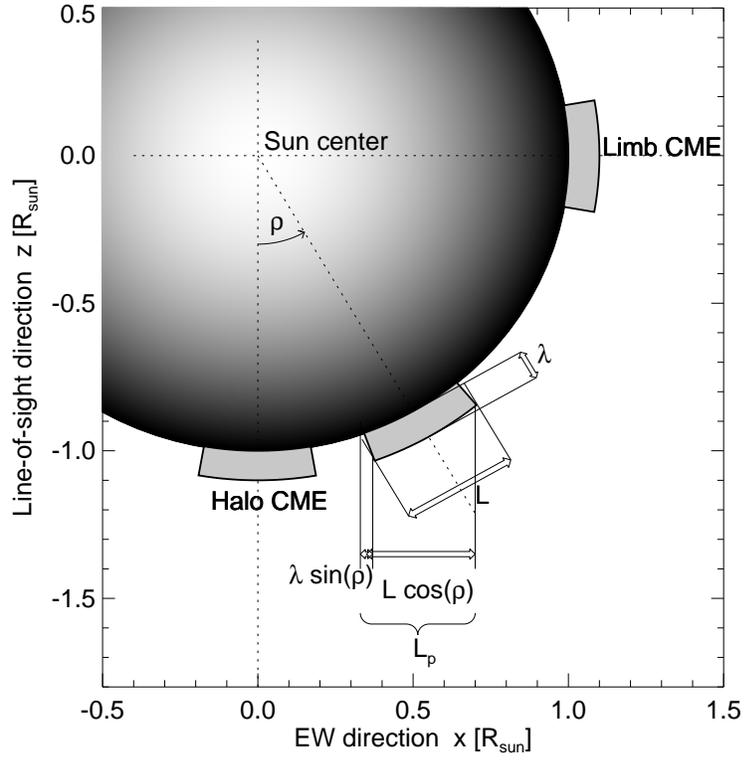}
\caption{Geometric relationship of a CME-associated EUV dimming region,
between the projected length scale $L_p = \lambda \sin{(\rho)} + L \cos{(\rho)}$ 
and the unprojected length scale $L$, the density scale height $\lambda$, and the 
aspect angle $\rho$, which approximately corresponds to the heliographic 
longitude for a CME at low latitude. Note that the extreme values are 
$L_p=L$ for a halo CME, and $L_p=\lambda$ for a limb CME, respectively.}
\end{figure}

\begin{figure}
\plotone{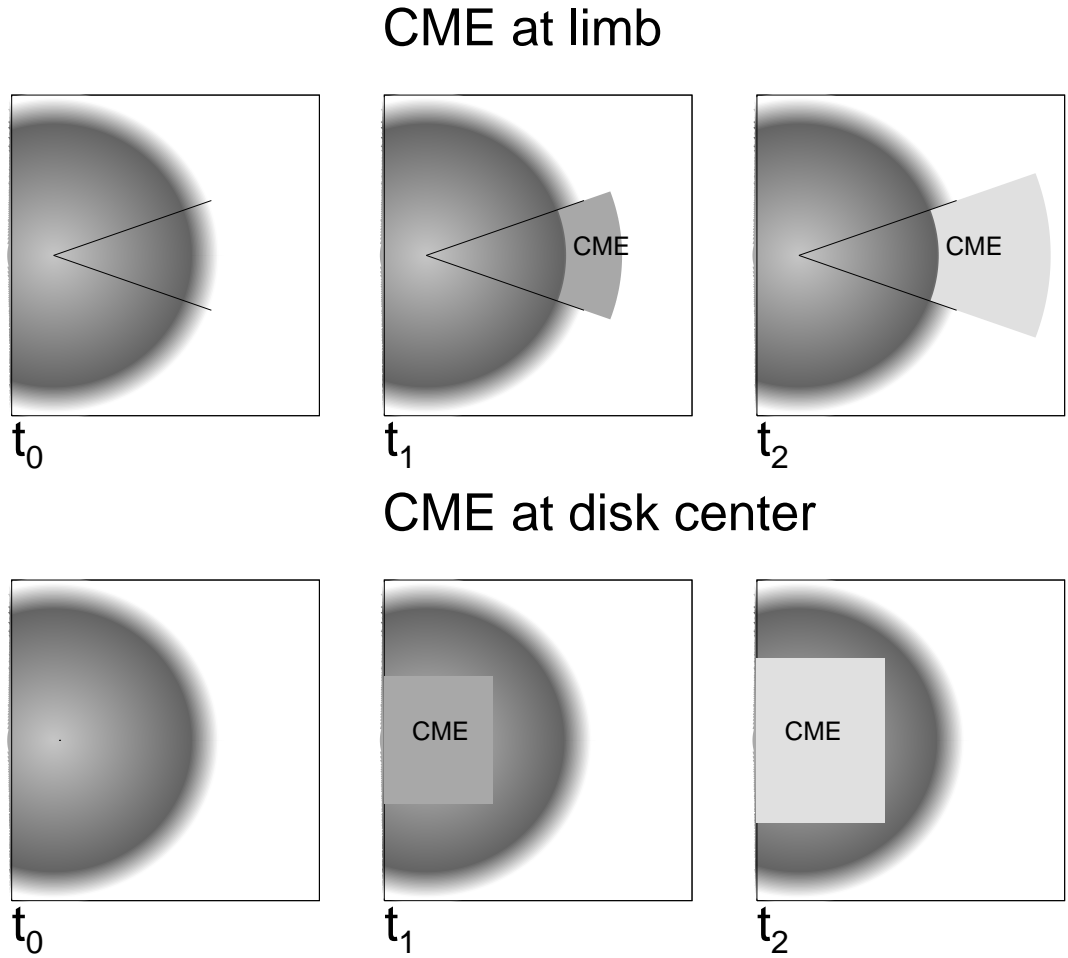}
\caption{Geometric model of a CME at the limb (top row) and a halo
CME at disk center (bottom row), shown for 3 subsequent time steps each.
The CME footpoint area can have an arbitrarily complex shape, but is 
rendered here with a gravitationally stratified equivalent square area.
The grey levels from dark to light indicate a decreasing emission measure
(or EUV dimming) from the CME volume, caused by the rarefaction of the CME
plasma when the CME expands in size.}
\end{figure}

\begin{figure}
\plotone{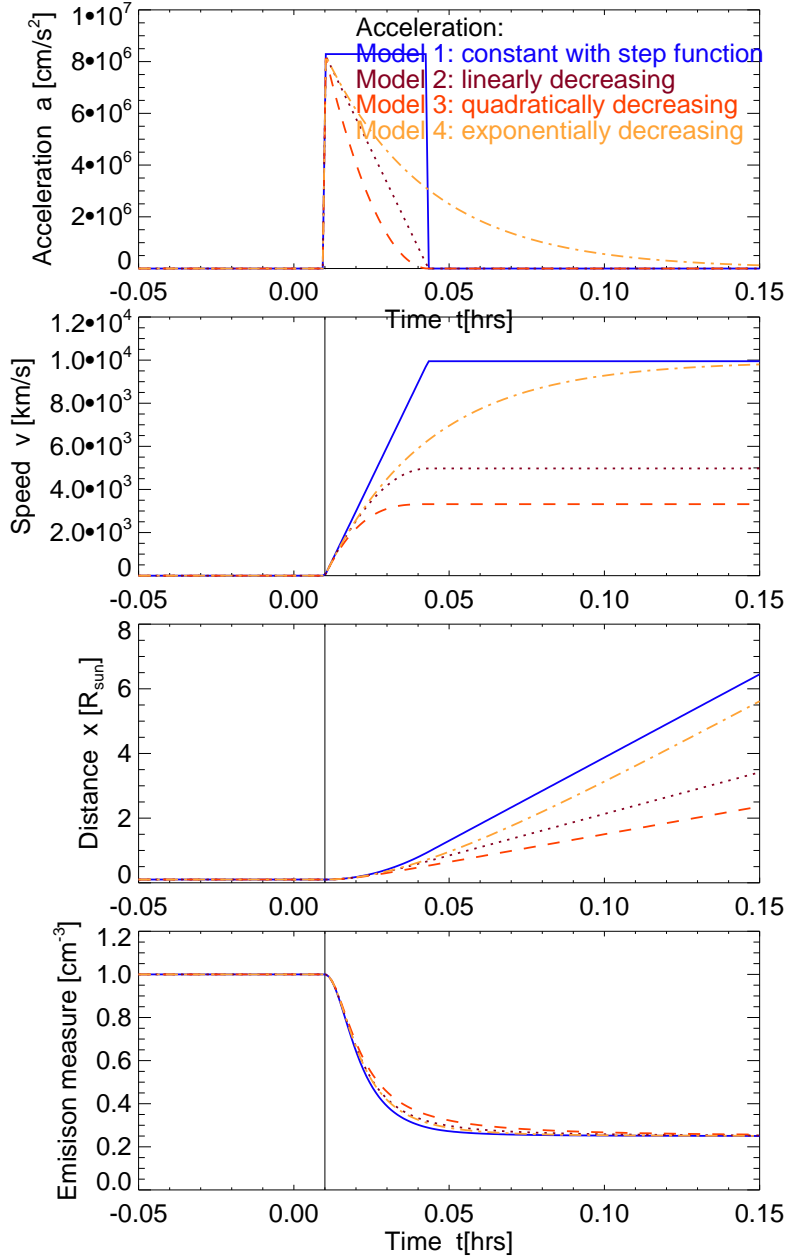}
\caption{Four models of the acceleration profile $a(t)$ (top panel),
the velocity profile $v(t)$ (second panel), the height-time profile $x(t)$
(third panel), and the emission measure profile $EM(t)$ (bottom panel). 
The start of the EUV dimming is indicated with a vertical line.}
\end{figure}

\begin{figure}
\plotone{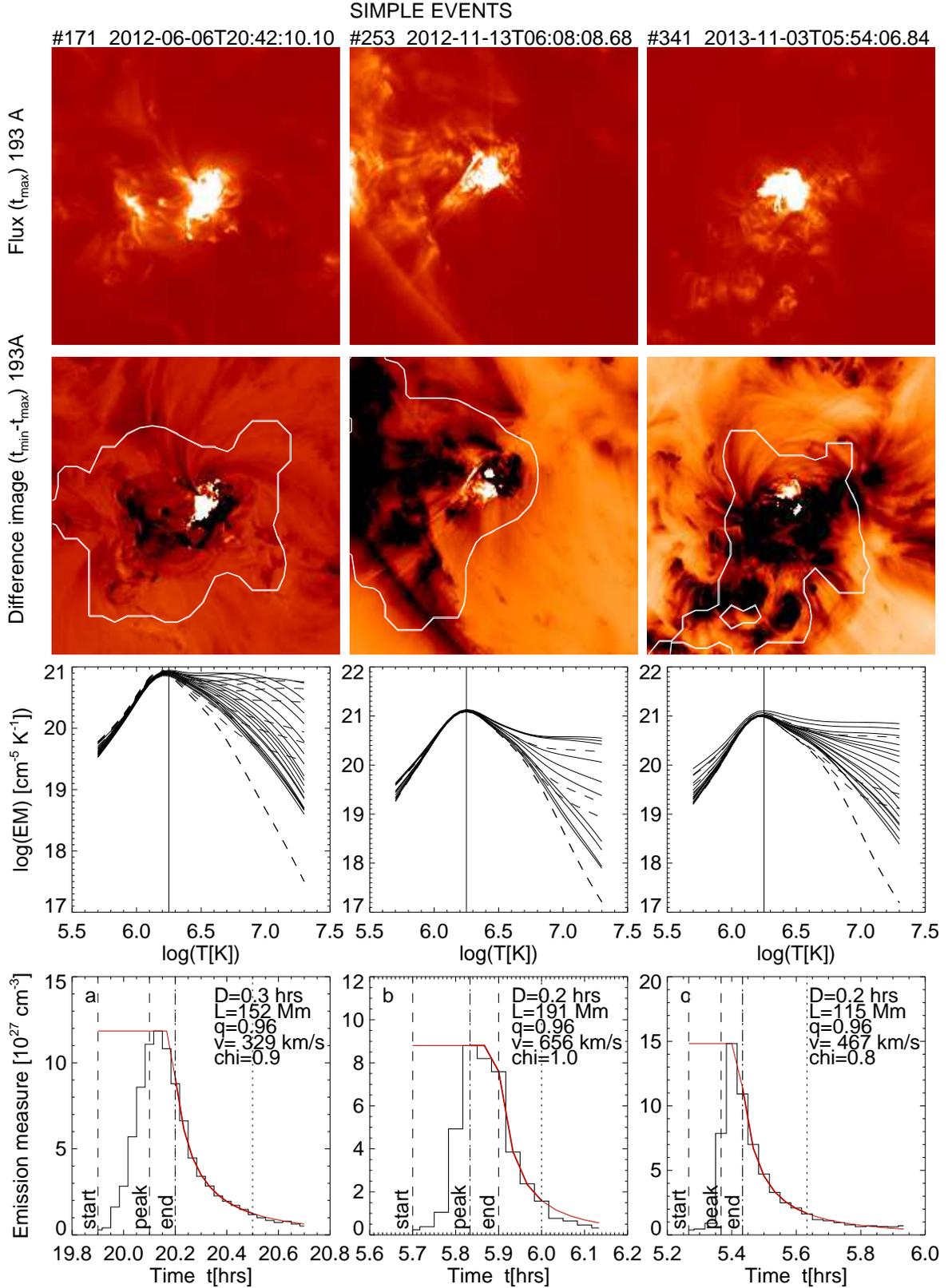}
\caption{The AIA 193 \ang\ image
taken at the peak time of the emission measure (top row), a time difference
image between the emission measure maximum time $t_{\mathrm{max}}$ and following
minimum time $t_{\mathrm{min}}$ (second row), the differential emission measure
distributions for each time step, marked with dashed curves at $t < t_{\mathrm{max}}$
and solid curves at $t > t_{\mathrm{max}}$ (third row), and the EUV emission measure
dimming curve (histogram) with fits of the radial adiabatic expansion model 
(red curve; bottom row) are shown for 3 simple events (\#171, 253, 341).
The area of EUV dimming is indicated with a white contour in the second row.}
\end{figure}

\begin{figure}
\plotone{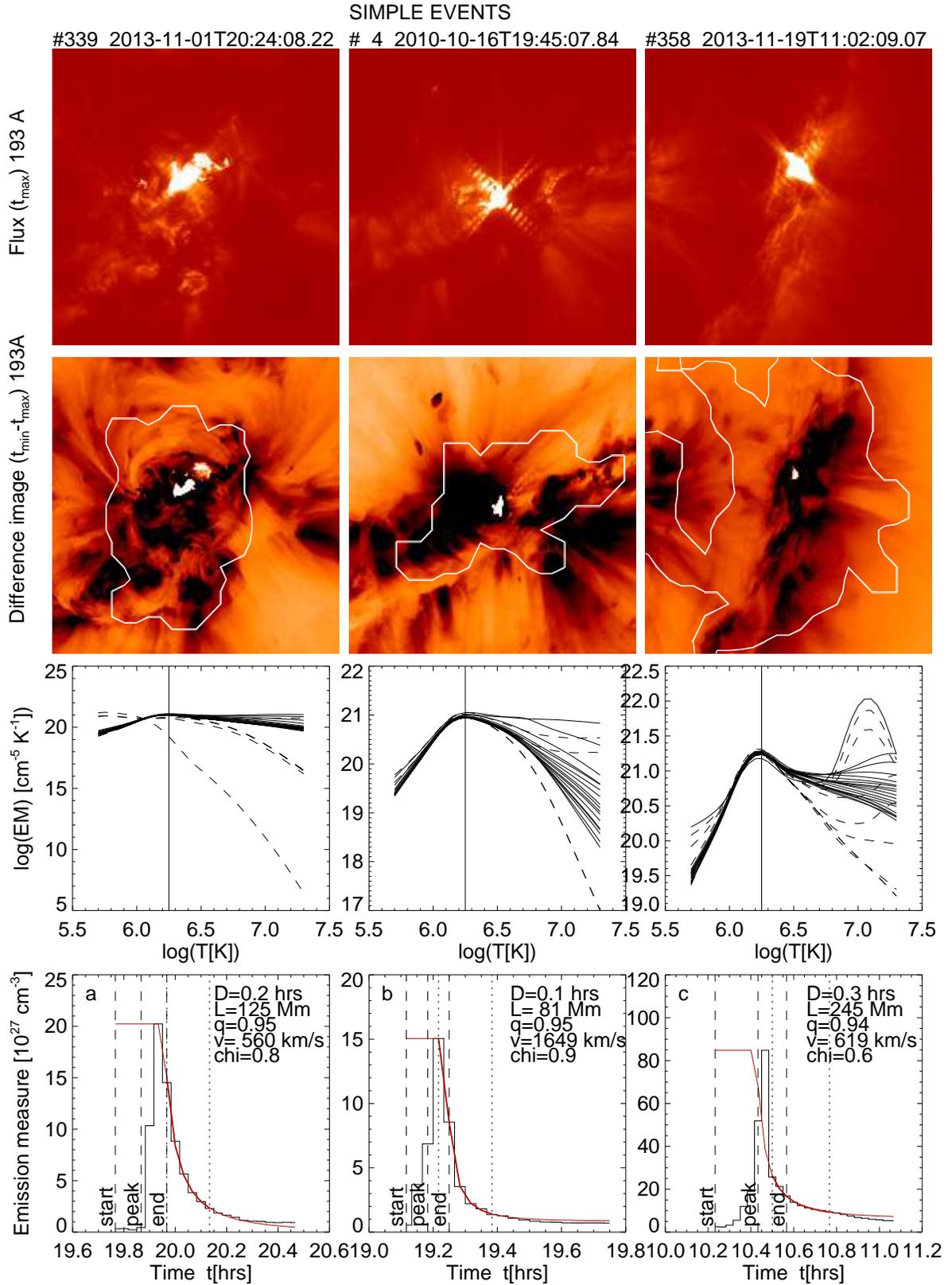}
\caption{Similar presentation as Fig.~5, for 3 simple events (\#339, 4, 358).}
\end{figure}

\begin{figure}
\plotone{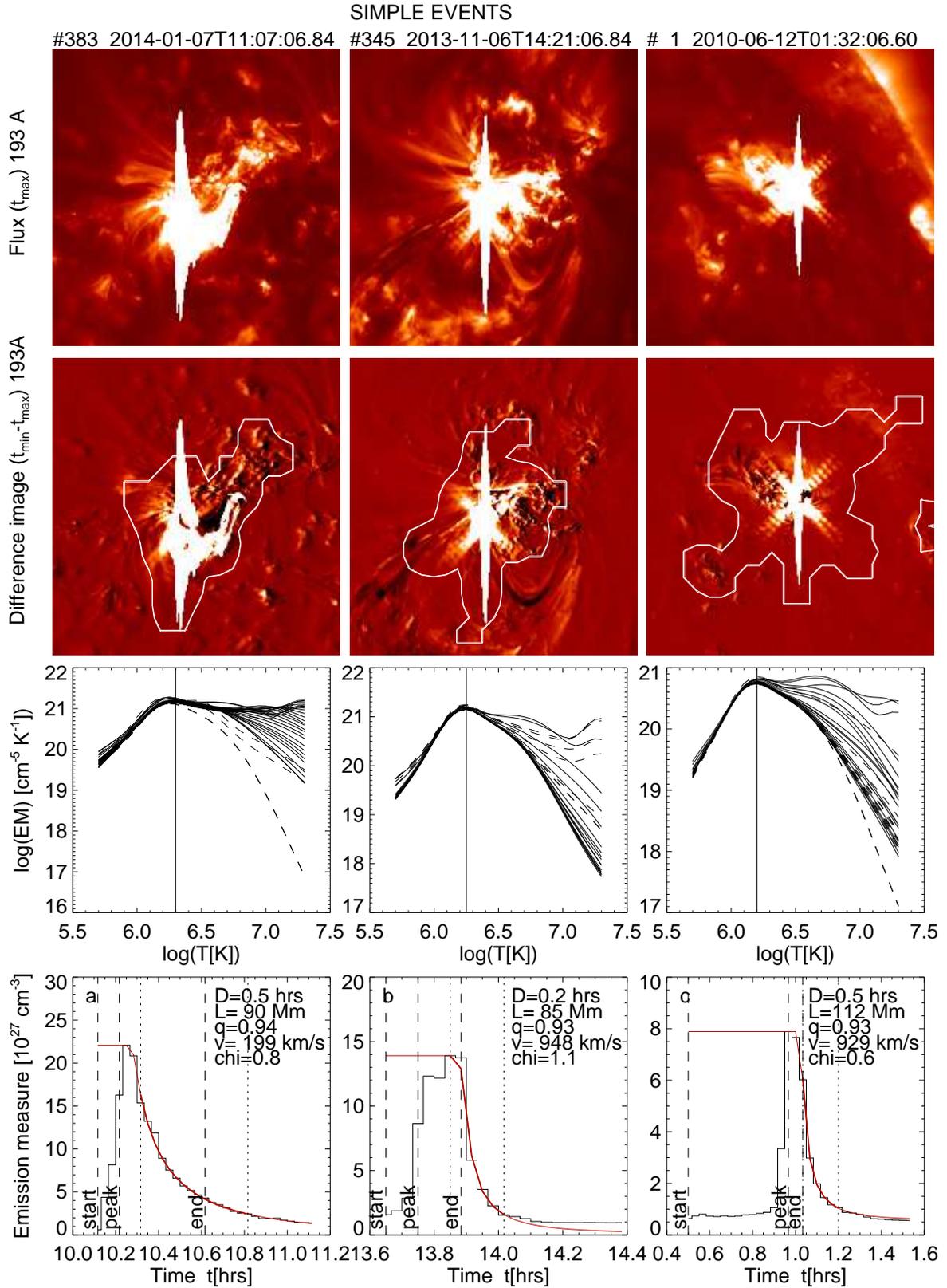}
\caption{Similar presentation as Fig.~5, for 3 simple events (\#383, 345, 1).}
\end{figure}

\begin{figure}
\plotone{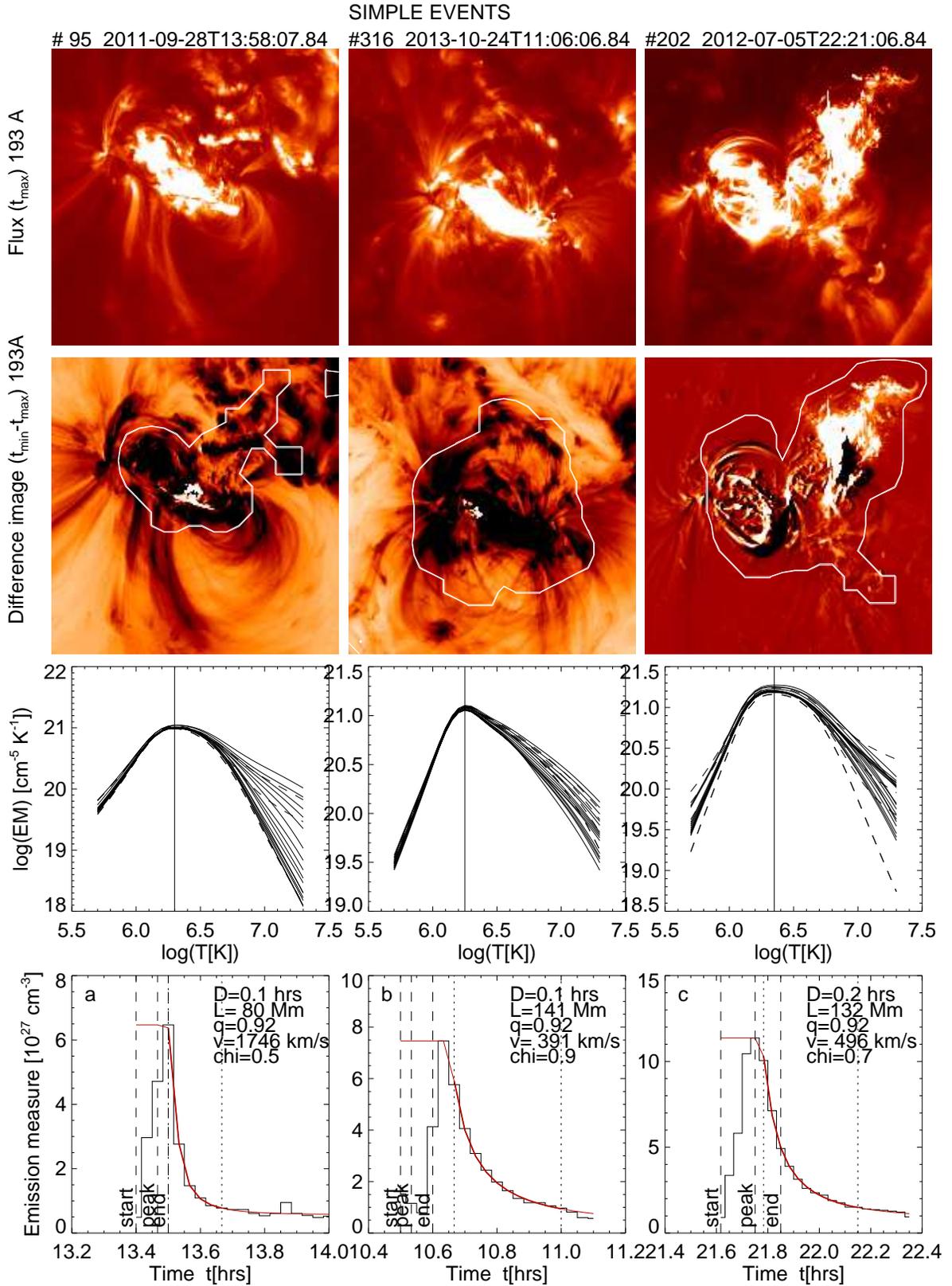}
\caption{Similar presentation as Fig.~5, for 3 simple events (\#95, 316, 202).}
\end{figure}

\begin{figure}
\plotone{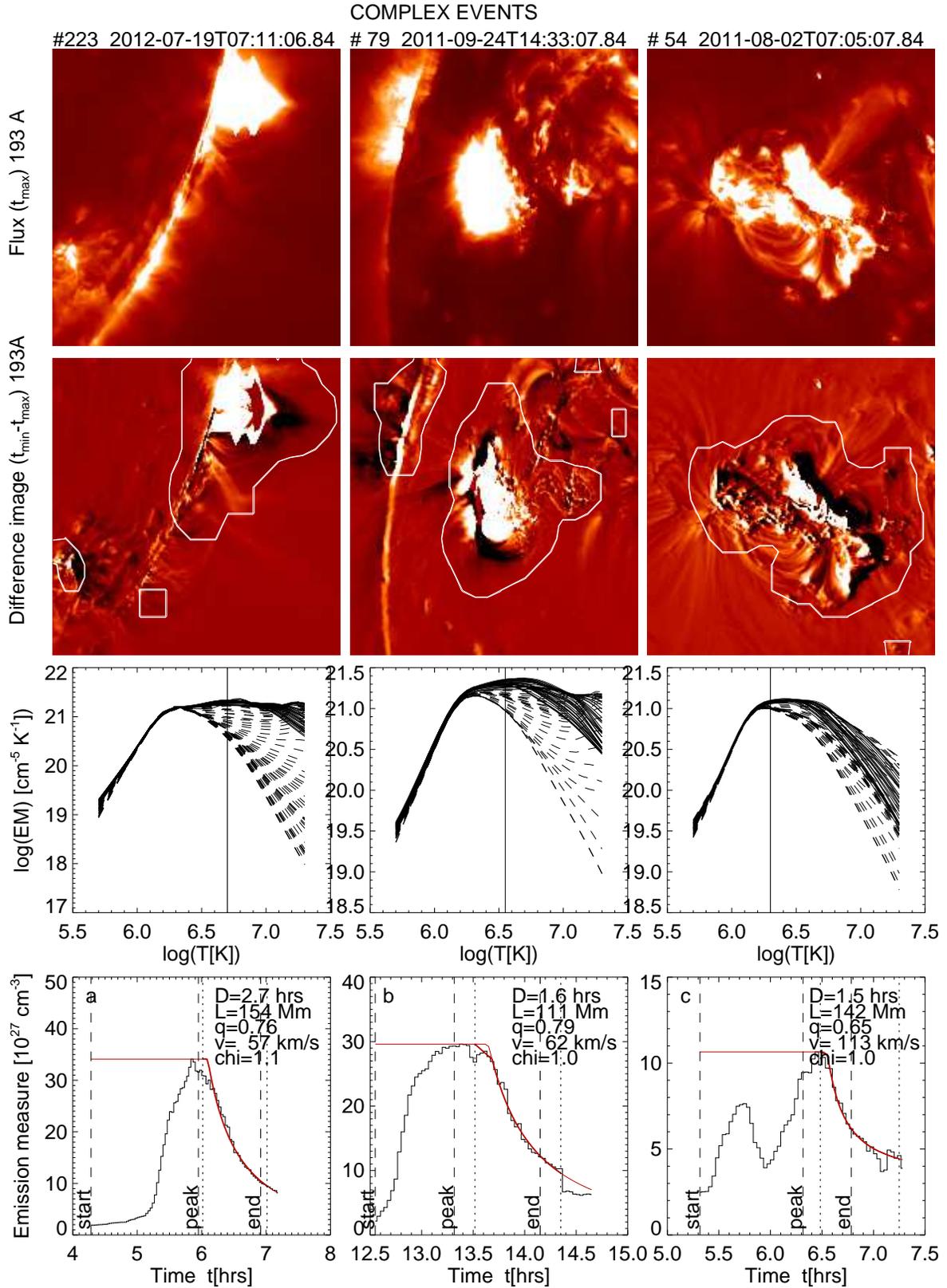}
\caption{Similar presentation as Fig.~5, for 3 complex events (\#223, 79, 54).}
\end{figure}

\begin{figure}
\plotone{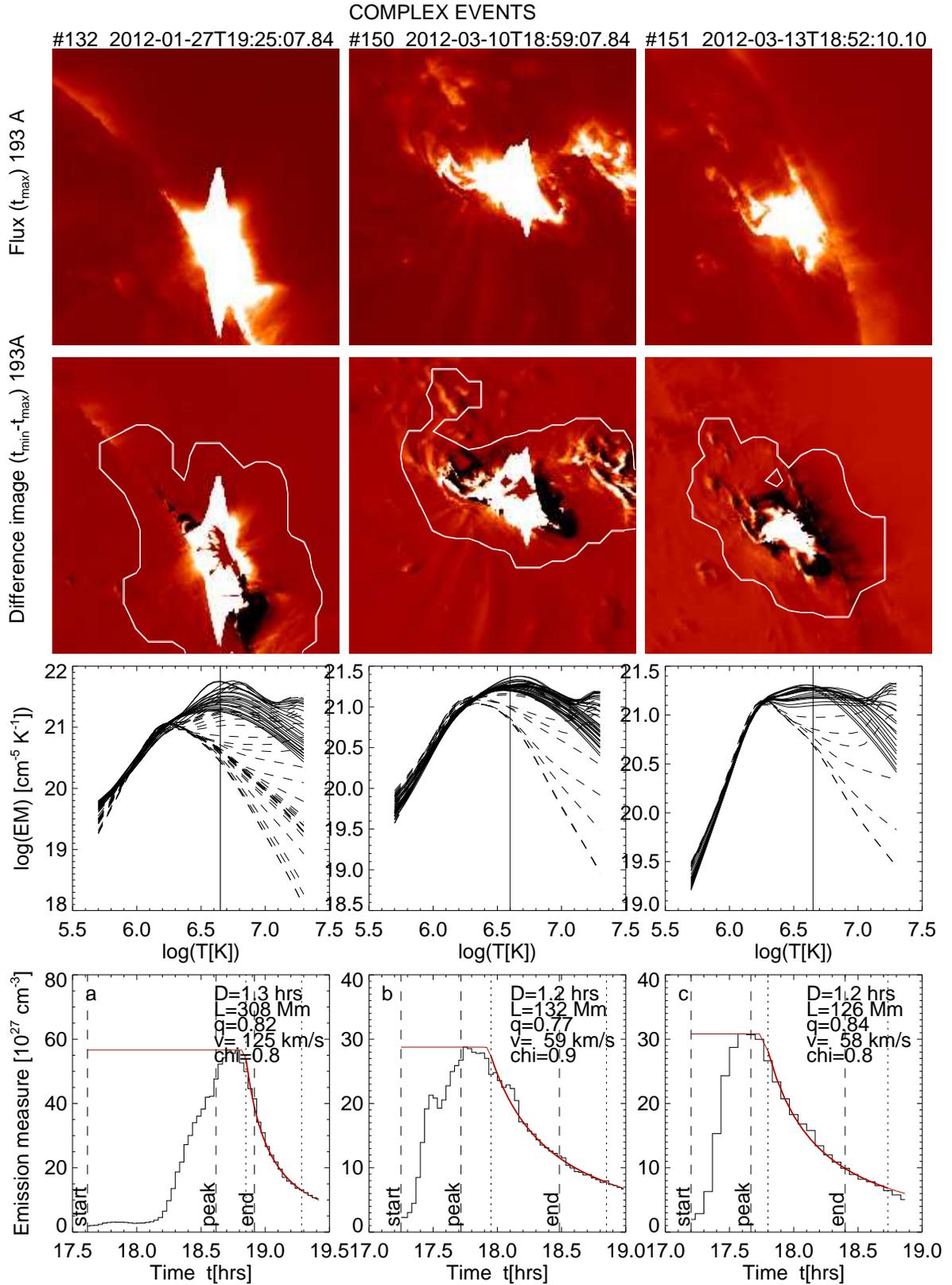}
\caption{Similar presentation as Fig.~5, for 3 complex events (\#132, 150, 151).}
\end{figure}

\begin{figure}
\plotone{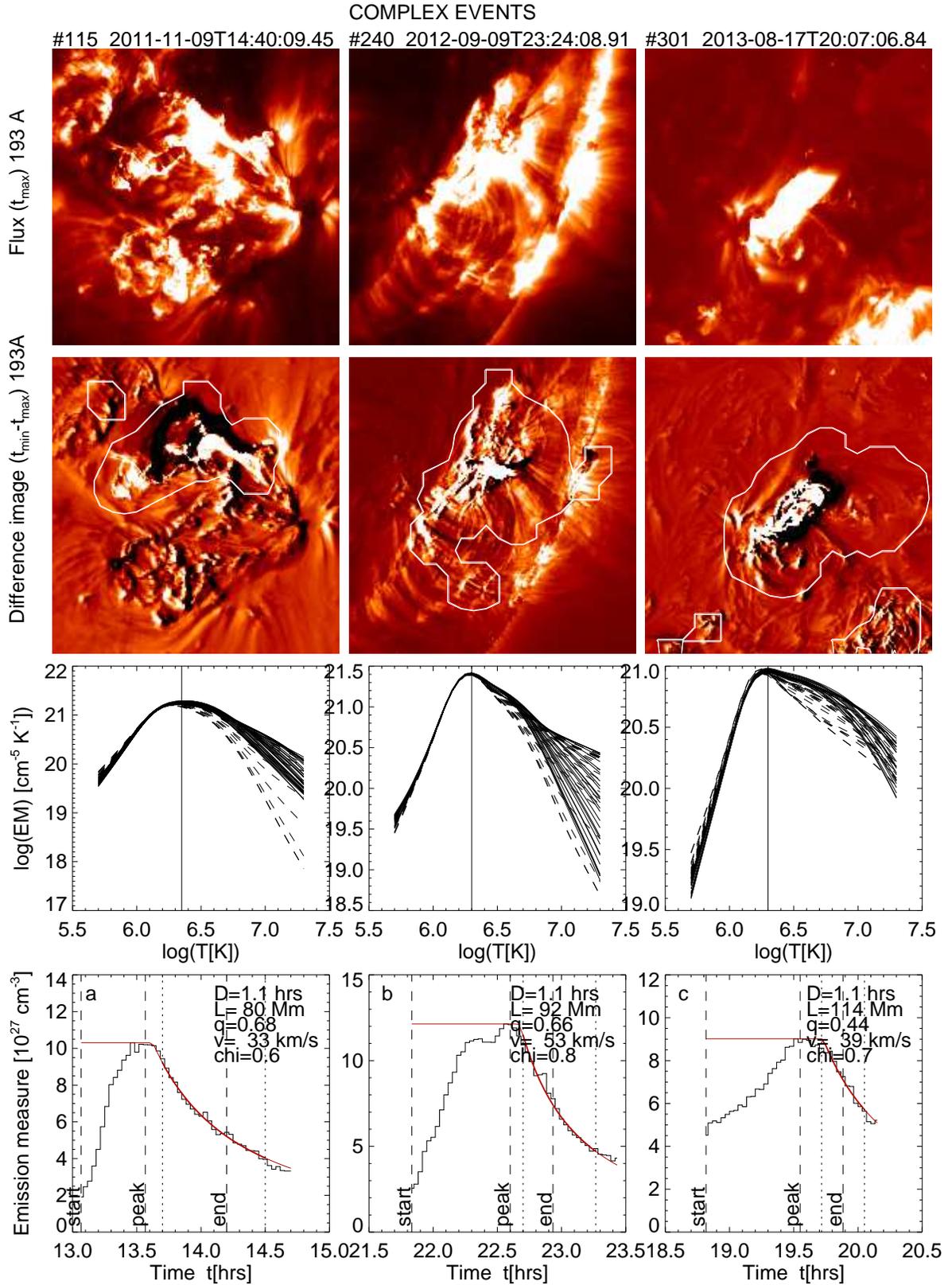}
\caption{Similar presentation as Fig.~5, for 3 complex events (\#115, 240, 301).}
\end{figure}

\begin{figure}
\plotone{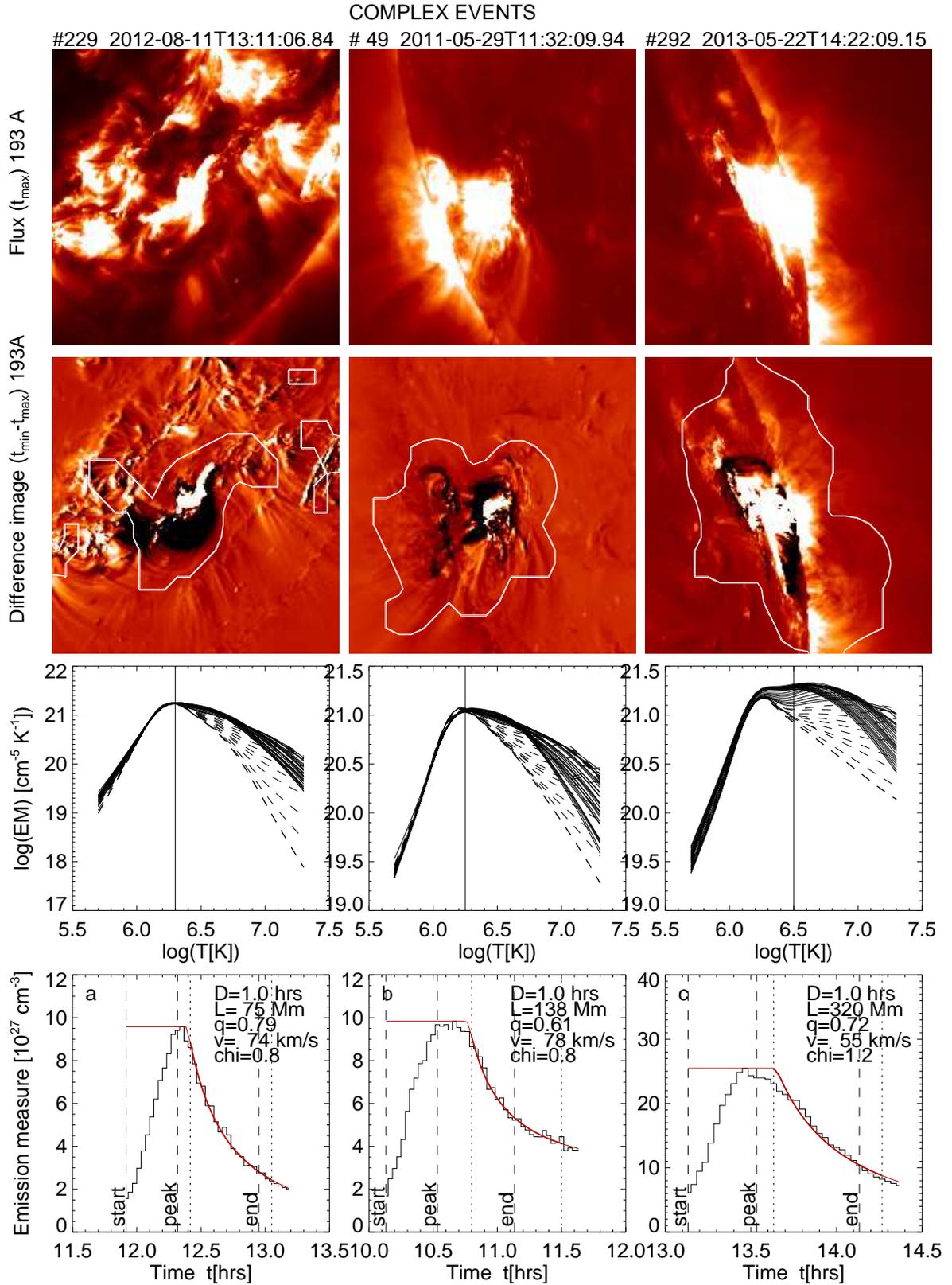}
\caption{Similar presentation as Fig.~5, for 3 complex events (\#229, 49, 292).}
\end{figure}

\begin{figure}
\plotone{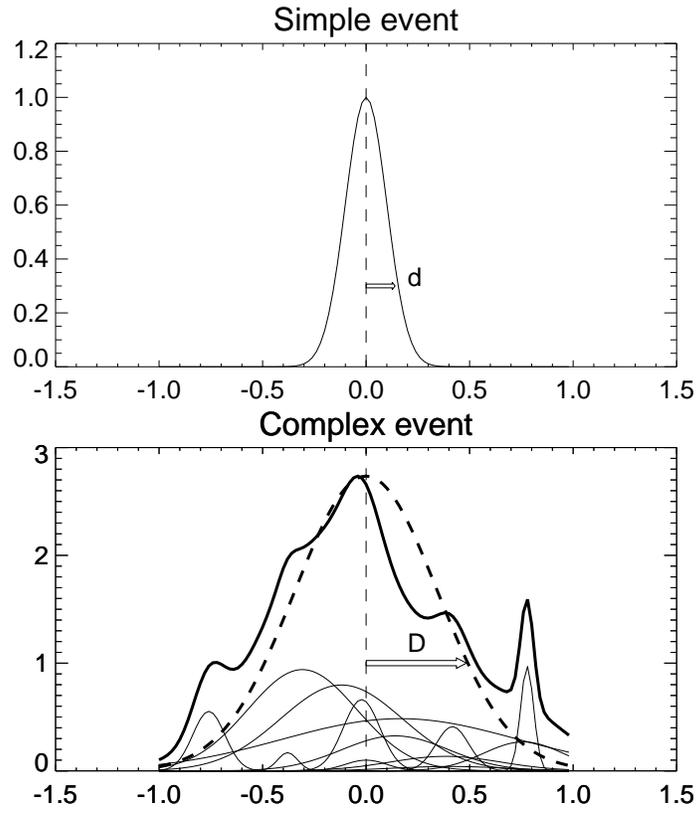}
\caption{{\sl Top panel:} Simple event with a Gaussian time profile and (half width)
duration $d$. {\sl Bottom panel:} Complex event (thick curve) consisting of a
random superposition of $N=12$ single Gaussians (thin curves), which have a predicted 
(half width) duration of $D = d \sqrt{N}$ (thick dashed curve).}
\end{figure}

\begin{figure}
\plotone{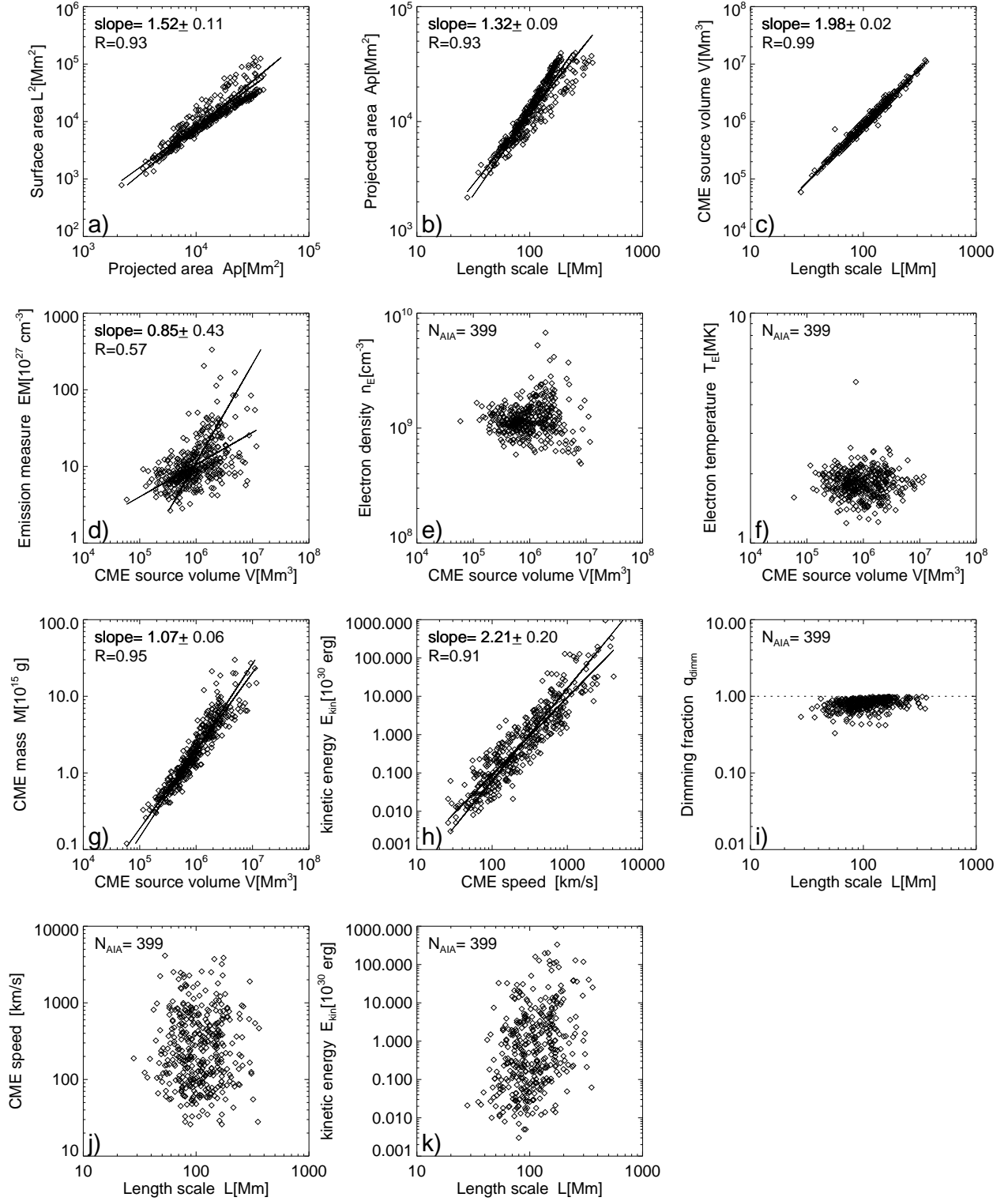}
\caption{Scatterplots of geometric parameters (projected area, CME surface area,
CME source volume), the CME emission measure, the CME electron density and
CME electron temperature (in the preflare phase), the CME mass, CME speed, 
CME kinetic energy, and dimming fraction.}
\end{figure} 

\begin{figure}
\plotone{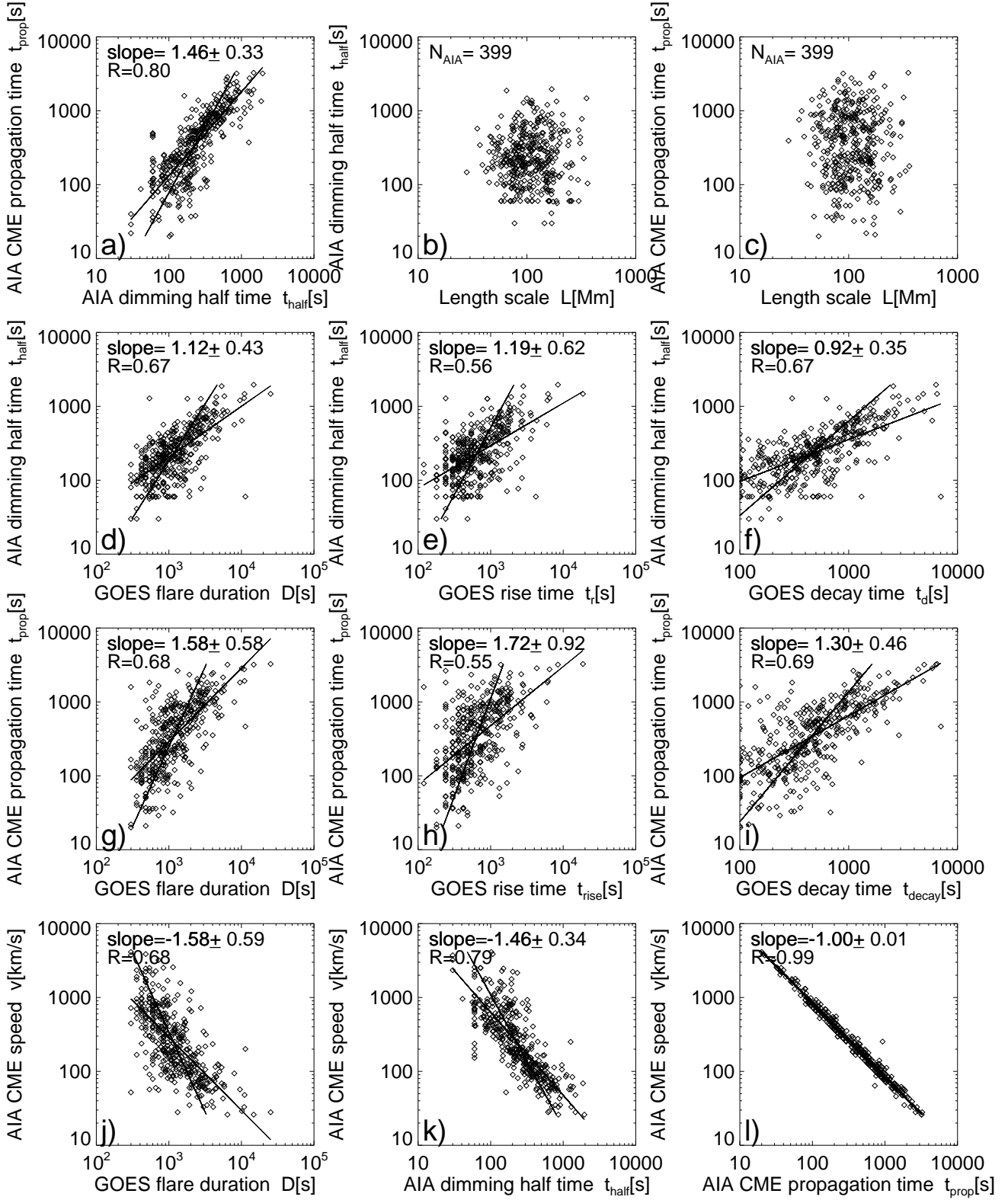}
\caption{Scatterplots of GOES and AIA temporal parameters of CMEs.
Two linear regression fits, $y(x)$ and $x(y)$, are indicated with solid black lines. }
\end{figure}

\begin{figure}
\plotone{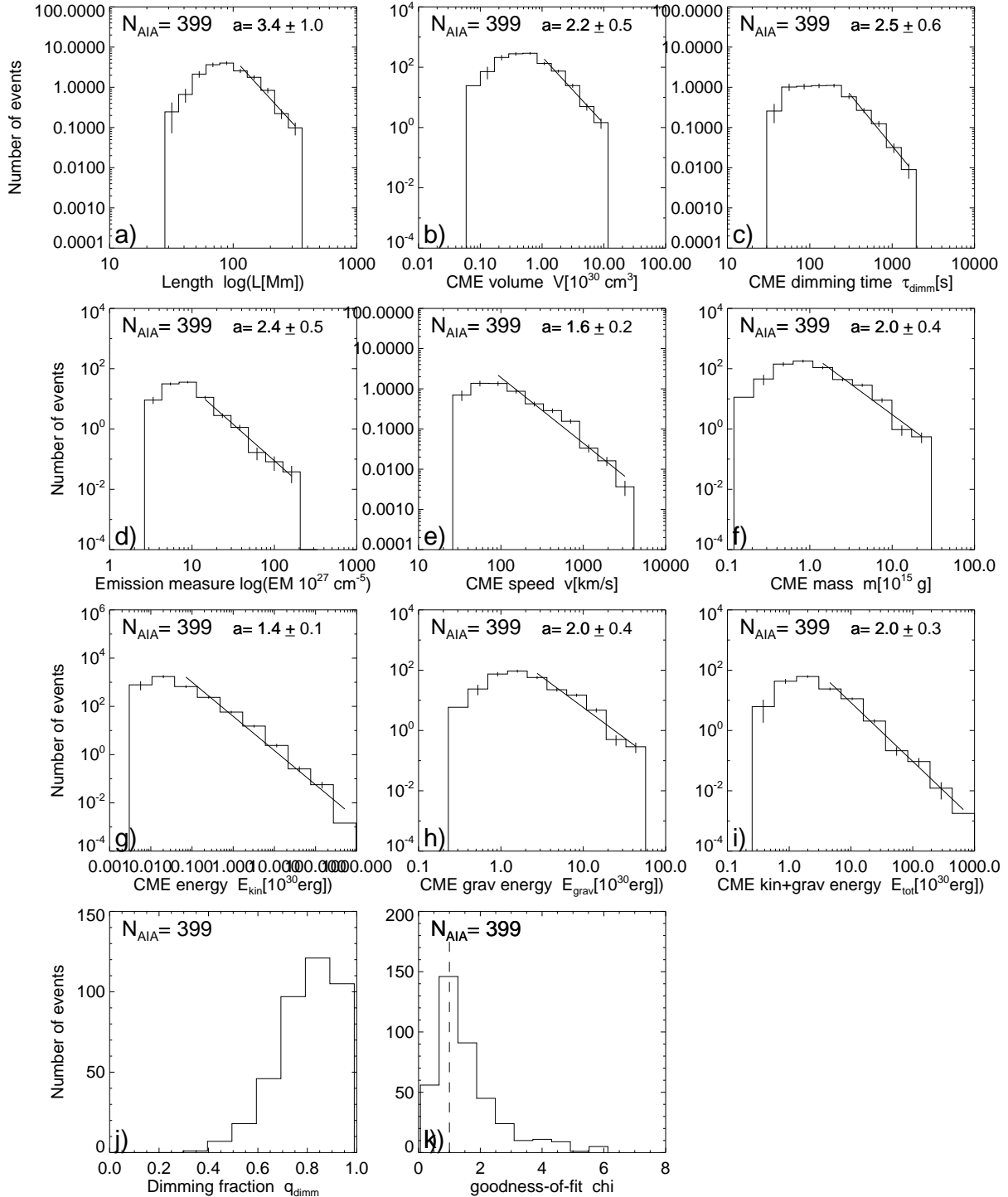}
\caption{Log-log histograms of various physical parameters measured from the
399 CME events based on AIA/SDO data. Power law fits are applied on the
right-side tails of the distributions.}
\end{figure}
\clearpage

\begin{figure}
\plotone{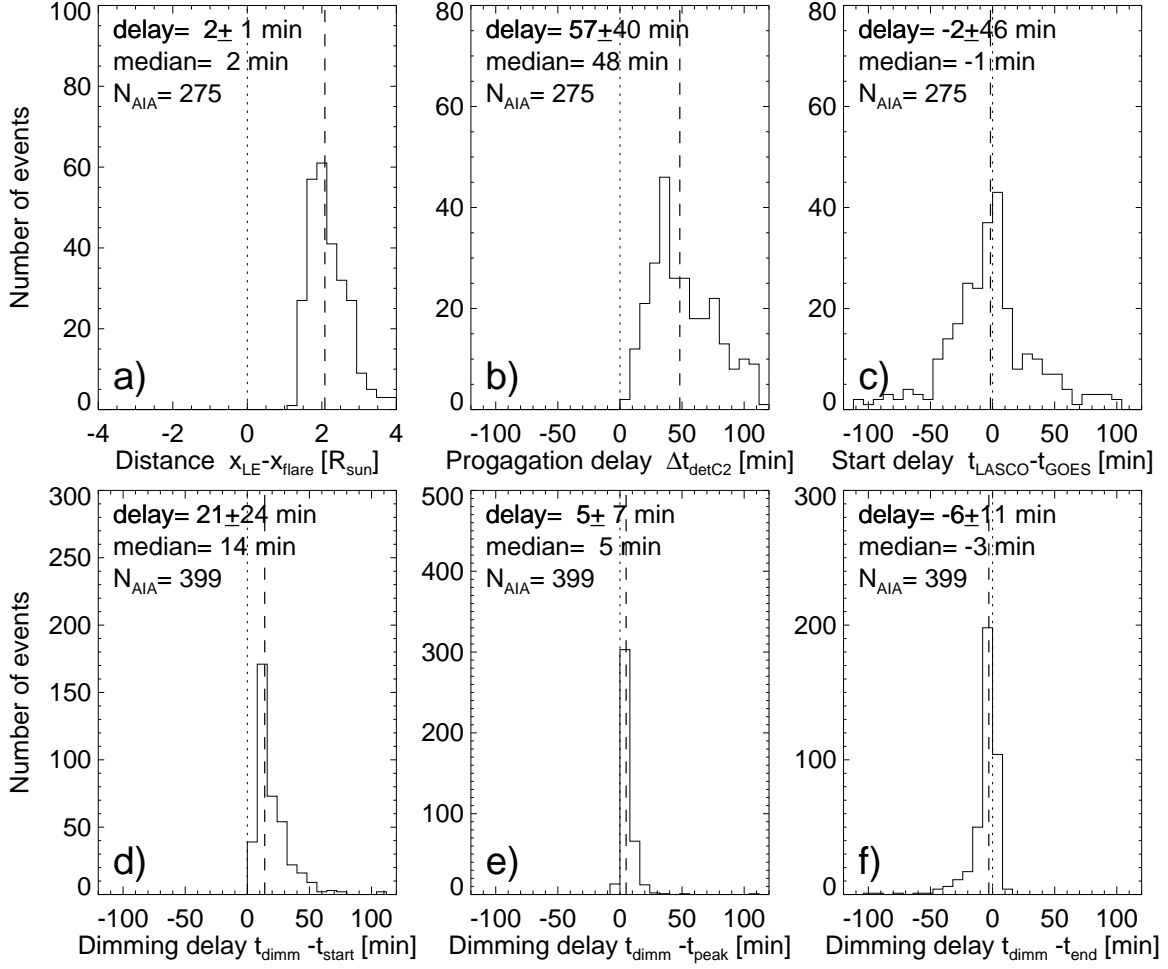}
\caption{The distance from the coronal flare site to the LASCO/C2
detection site (a), the LASCO/C2 detection delay (due to propagation)
with respect to the predicted
CME onset time (b), and with respect to the GOES flare start time (c)
are shown (upper panels). Distributions of time delays of EUV dimming times 
$t_{\mathrm{dimm}}$ with respect to the GOES start time (d), peak time (e), and 
end time (f) are shown in the lower panels. }
\end{figure}

\begin{figure}
\plotone{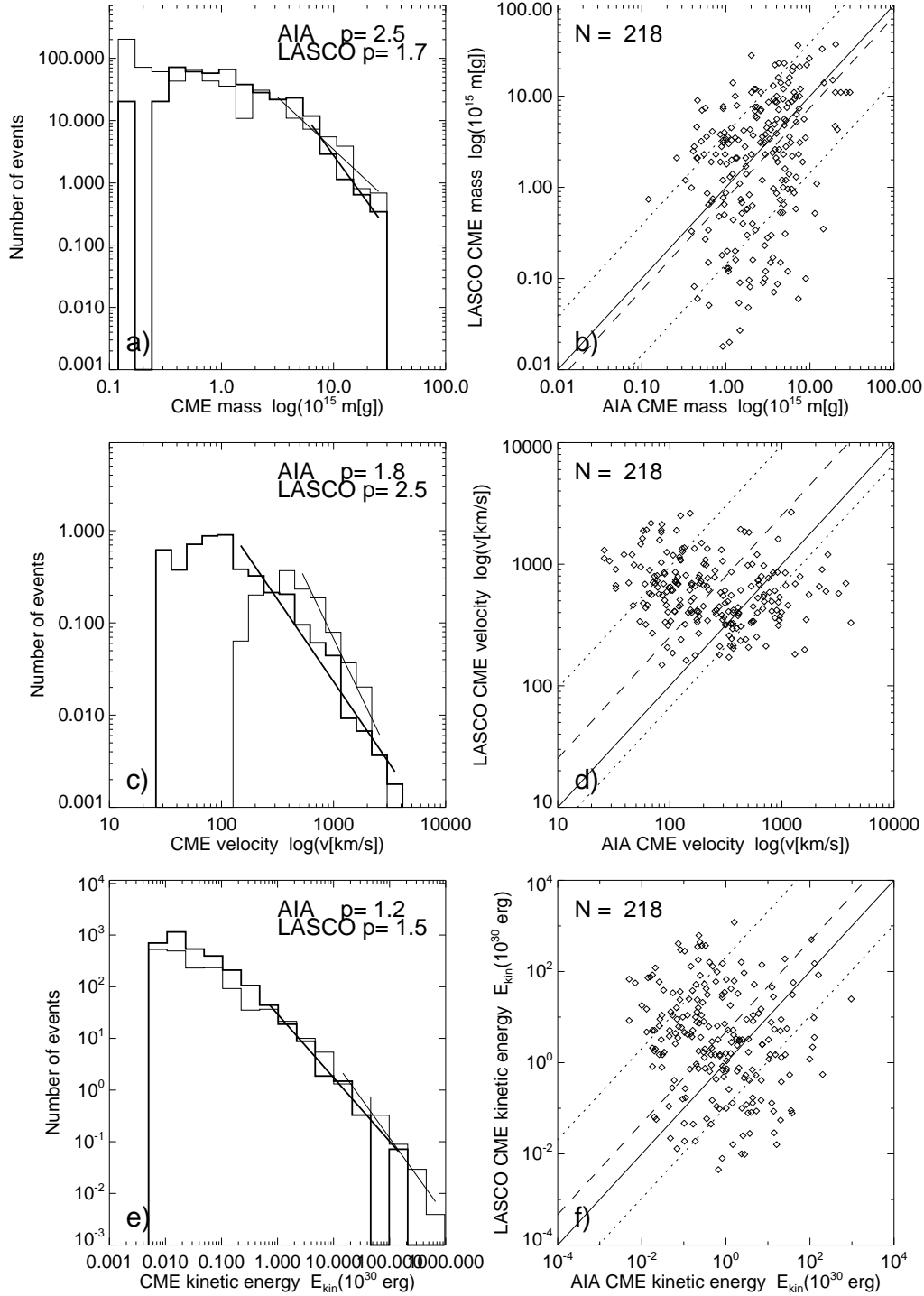}
\caption{Comparison of of CME masses (top panels), CME velocities
(middle panels), and CME kinetic energies (bottom panels) between AIA and LASCO
data sets, in form of (log-log) size distributions (left panels) and
scatterplots (right panels). The diagonal solid line indicates identity, 
along with the means (dashed line) and standard deviations (dotted lines) 
of the logarithmic ratios.}
\end{figure}

\begin{figure}
\plotone{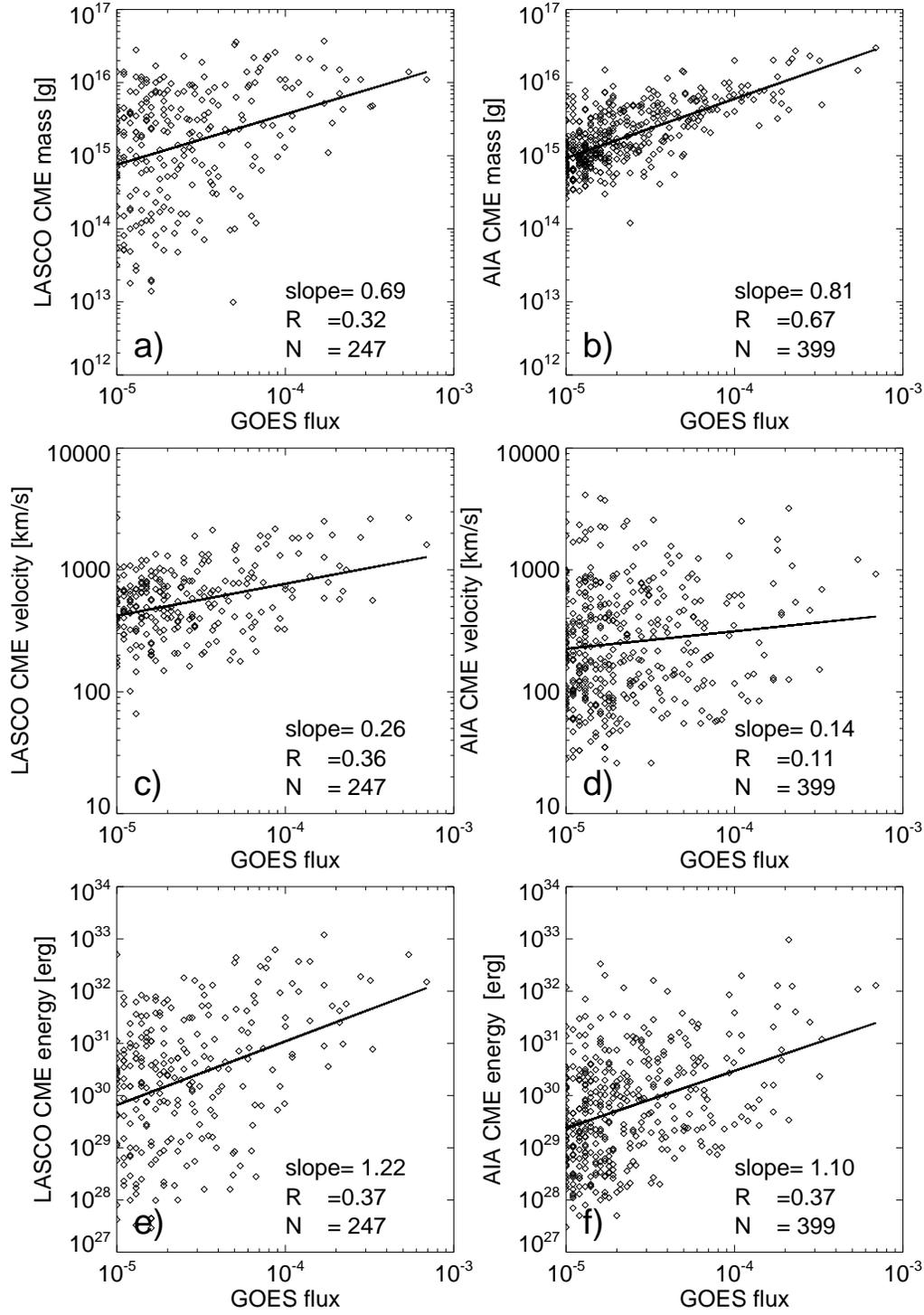}
\caption{Scatterplot of LASCO/C2 (left panels) and AIA (right panels) CME masses
(top panels), CME velocities (middle panels), and CME kinetic energies (right
panels) versus the GOES flux.}
\end{figure}

\begin{figure}
\plotone{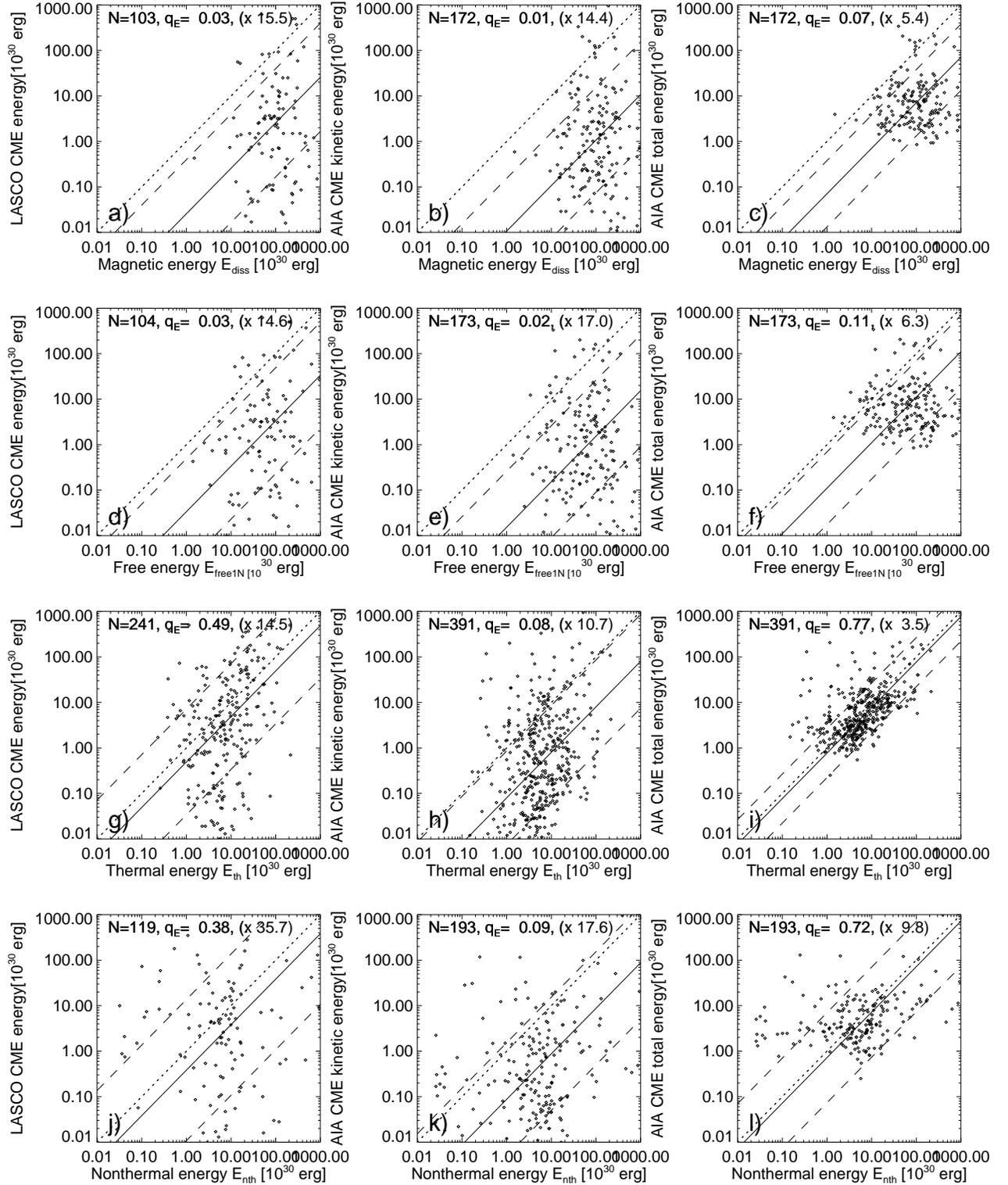}
\caption{Scatterplot of CME kinetic energies versus the dissipated magnetic energy (top row),
the free (magnetic) energy (second row), the multi-thermal energy (third row),
and the nonthermal energy (bottom row), for both LASCO/C2 (left column) and
AIA observations (middle column). The total (kinetic and gravitational) energies
are shown in the right column. Equality (dotted diagonal line), the mean 
(dashed line) and standard deviation (dotted line) of the energy ratios are
indicated.}
\end{figure}

\end{document}